\documentclass[amsmath,amssymb,10pt,aps,twocolumn]{revtex4-1} 
 \usepackage[colorlinks=true]{hyperref}
 \usepackage{lipsum}
 \usepackage{graphicx}
 \usepackage{textcomp}
\usepackage[dvipsnames]{xcolor}
\usepackage{latexsym}
\usepackage{amsmath}
\usepackage{amsthm}
\usepackage{amssymb}
\usepackage{epstopdf} 
\usepackage{enumerate}
\usepackage{setspace} 
\usepackage{dcolumn}
\usepackage{bm}
\usepackage{setspace} 
\usepackage{slashed}
\usepackage{color}
\usepackage{xcolor}
\usepackage{youngtab}
\usepackage{tikz}
\usepackage{tikz-3dplot}
\usepackage{braket}
\usepackage{mhchem}
\usepackage{float}
\usetikzlibrary{shapes,snakes,arrows,chains,matrix,positioning,scopes,calc}
\usetikzlibrary{decorations.markings}

\usepackage[tabbotcap]{subfigure}

\allowdisplaybreaks
 
\begin{document}

\title{
Non-Fermi liquid behavior and quantum criticality in cubic heavy fermion systems with non-Kramers multipolar local moments
}

\author{SangEun Han}
\thanks{These authors contributed equally to this work.}
\affiliation{Department of Physics, University of Toronto, Toronto, Ontario M5S 1A7, Canada}

\author{Daniel J. Schultz}
\thanks{These authors contributed equally to this work.}
\affiliation{Department of Physics, University of Toronto, Toronto, Ontario M5S 1A7, Canada}

\author{Yong Baek Kim}
\affiliation{Department of Physics, University of Toronto, Toronto, Ontario M5S 1A7, Canada}
\affiliation{School of Physics, Korea Institute for Advanced Study, Seoul 02455, Korea}
\date{\today}

\begin{abstract}
Notable non-Fermi liquid and quantum critical behaviors are observed in rare-earth metallic systems with non-Kramers local moments supporting a number of different multipolar moments. A prominent example is Pr(Ti,V)$_2$Al$_{20}$, where the non-Kramers doublet of the Pr$^{3+}$ ion allows quadrupolar and octupolar moments, but lacks a dipolar moment. Previous theoretical studies show that a single impurity Kondo problem with such an unusual local moment leads to novel non-Fermi liquid states. In this work, we investigate possible quantum critical behaviors arising from the competition between non-Fermi liquid states and multipolar-ordered phases induced by the RKKY interaction. We consider a local version of the corresponding Kondo lattice model, namely the Bose-Fermi Kondo model. Here, the multipolar local moments are coupled to fermionic and bosonic bath degrees of freedom representing the multipolar Kondo effect and RKKY interactions. Using a perturbative renormalization group (RG) study up to two loop order, we find critical points between non-Fermi liquid Kondo fixed points and a quadrupolar ordered fixed point. The critical points describe quantum critical behaviors at the corresponding phase transitions and can be distinguished by higher order corrections in the octupolar susceptibility that can be measured by ultrasound experiments. Our results imply the existence of a rich expansion of the phases and quantum critical behaviors in multipolar heavy fermion systems. 
\end{abstract}

\maketitle

\section{Introduction}
Due to strong spin-orbit coupling and crystal electric field effects, $f$-electrons in rare-earth metallic systems may form higher-rank multipolar moments \cite{Kusunose2008b, Kuramoto2009a}. Recent studies on the relevant materials have shown that the interplay between the multipolar moments and itinerant conduction electrons can lead to a variety of exotic quantum phases of matter including multipolar ordered phases, strange metallic behavior, and unconventional superconductivity \cite{Martelli2019, Rosenberg2019c, White2015, Saxena2000, Kohori2000, Izawa2001, Aoki2003, Bauer2004,Cox1987a, Jiao2015, Kratochvilova2015, Custers2010a, Falkowski2014, Custers2012, Cameron2016, Rylands2022}; this interplay also provides a setting for quantum critical behavior \cite{Stewart2001, Lohneysen1994, Abrahams2012}. In contrast to the archetypal heavy fermion systems with dipolar local moments \cite{Irkhin2016, Irkhin2017}, such materials are described by a multipolar Kondo lattice model, with the Kondo interactions and subsequent RKKY interactions taking very unusual forms \cite{Iizuka2020, Lai2018, Kuzmenko2018}. The \ce{Pr}-based caged cubic compounds \ce{Pr(Ti,V)2Al20} are a particularly exciting realization of a multipolar Kondo lattice. In this class of materials, the \ce{Pr^{3+}} ions rest at the centre of a tetrahedral cage, and contribute a local $4f^2$ moment. The moment forms a non-Kramers doublet in the presence of the crystal field, and lacks any dipolar moment; instead it only carries quadrupolar and octupolar moments. These moments provide a route a number of experimentally observed phenomena including a quadrupolar ordered phase, a multipolar Kondo effect related to non-Fermi liquid behavior (with resistivity $\rho\sim T^{1/2}$ for the case of \ce{PrV2Al20}), and the multipolar fluctuations responsible for the unconventional superconducting phase \cite{Onimaru2011a, Onimaru2016b, Onimaru2016c, Freyer2018a, Sato2012a, Araki2014, Matsushita2011, RameshKumar2016, Sakai2011b, Worl2019a, Matsubayashi2012c, Sakai2012a, Tsujimoto2014a, Fu2020a, Onimaru2012, Matsumoto2015, Onimaru2010, Shimura2015a, Nagashima2014a}.

Theoretical works to understand these different quantum phases have either focused on the properties deep inside of a phase \cite{Schultz2021b, Patri2020e, Patri2022, Patri2019d, Tsuruta2015, Tsuruta1999, Tsuruta2000a, Tsuruta2022, Patri2020d}, or have been Ginzburg-Landau studies of magnetic ordering phase transitions \cite{Lee2018e}. However, a full understanding of the quantum critical behavior which may arise due to the competition between multipolar RKKY fluctuations and the Kondo effect remains elusive. A simplified approach to this competition is the Bose-Fermi Kondo model, which serves as a local approximation of the full Kondo lattice \cite{Si1996,Smith1999,Smith2000,Si2001, Si2003, Sengupta2000, Zarand2002, Kirchner2005, Zhu2002, Han2021}. Within this scheme, the Kondo coupling between the local moment and conduction electrons is taken into account explicitly, but the RKKY interaction is replaced by a dynamical bosonic bath with the density of states $\sim|\omega|^{1-\epsilon}$ acting at the impurity site. The main two kinds of phases admitted by renormalization group analyses of this model are a Kondo phase with large Fermi surface, wherein the local moments hybridize with the conduction electrons increasing the total density, and a magnetically ordered phase where the Kondo effect is destroyed and the conduction electrons are decoupled from the local moments \cite{Steglich2014, Friedemann2010, Friedemann2010a, Paschen2004, Coleman2001}.  The latter phase is to be contrasted against the Moriya-Hertz-Millis type (potentially multipolar) spin density wave instability of a heavy Fermi liquid \cite{Lohneysen2007}. Although other Bose-Fermi Kondo models studied with extended dynamical mean field theory may yield a spin-density wave type phase \cite{Si2001,Glossop2007,Zhu2007}, the results we will describe here fall into the Kondo destruction category. 

In this work, motivated by experiments on Pr-based heavy fermion systems, \ce{Pr(Ti,V)2Al20}, we consider the multipolar Bose-Fermi Kondo model in cubic systems as a simplified model for the multipolar Kondo lattice. Here, the Pr$^{3+}$ ions provide a non-Kramers doublet supporting quadrupolar and octupolar moments, which couple to conduction electrons. As shown below, the multipolar Kondo lattice model permits a mapping to a multipolar Bose-Fermi Kondo model, which can be constructed based on local point group symmetry. The fermionic Kondo problem without the bosonic bath or RKKY interaction was theoretically studied earlier and various non-Fermi liquid ground states were identified. These results may have some relevance to experiments in the dilute limit \cite{Yamane2018b, Yanagisawa2019b, Yanagisawa2019c, Patri2020e, Schultz2021b}, where non-Fermi liquid behaviors were observed. It is then natural to study the phase transition between such non-Fermi liquids and a multipolar ordered phase. 

\begin{figure}
    \centering
    \includegraphics[scale=1]{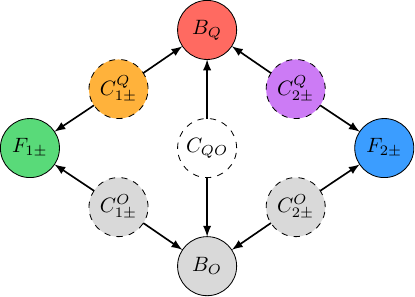}
    \caption{Schematic diagram for quantum phase transitions between non-Fermi liquid phases and multipolar ordered phases. $F_{1\pm}$ and $F_{2\pm}$ stand for the novel multipolar and 2-channel Kondo fixed points, respectively. $B_{Q}$ and $B_{O}$ stand for the quadrupolar and octupolar ordered fixed points, respectively, and $C_{QO}$ is a critical point between $B_{Q}$ and $B_{O}$. $C_{1\pm}^{Q(O)}$ and $C_{2\pm}^{Q(O)}$ are critical points between $F_{1\pm}$ and $B_{Q(O)}$, and $F_{2\pm}$ and $B_{Q(O)}$, respectively. Critical points have dashed lines, and stable fixed points have solid lines; gray circles are outside of the perturbative regime. The fixed point values are listed in Table~\ref{tab:fixed_point_suscep}.}
    \label{fig:destruction}
\end{figure}

In the multipolar Bose-Fermi Kondo model, we set the density of states of the quadrupolar ($Q$) and octupolar ($O$) bosonic baths to be $|\omega|^{1-\epsilon_i} (i = Q, O)$ and perform a perturbative renormalization group (RG) analysis based on an $\epsilon$-expansion to order $\epsilon^2$ to study the zero temperature phase diagram of the model. Similar to previous work, we find that there are two non-Fermi liquid phases \cite{Patri2020e, Patri2020d} in the fermion Kondo part of the model; one is a two-channel Kondo non-Fermi liquid, and the other is a novel non-Fermi liquid phase, not simply classifiable into any multichannel-type model. These phases, upon tuning the Kondo and bosonic bath couplings, can pass through quantum critical points to both arrive at a quadrupolar ordered phase, as presented in the schematic diagram (Fig.~\ref{fig:destruction}) and in the RG flow diagrams (Figs.~\ref{fig:phase1}-\ref{fig:phase2}).
The transition from the non-Fermi liquid phases to the quadrupolar ordered phase is accompanied by the destruction of the Kondo effect such that the Kondo coupling flows to zero in the quadrupolar ordered phases, representing a small Fermi surface state \cite{Paschen2004, Friedemann2010a}. 

To distinguish the critical points and non-Fermi liquid phases experimentally, we compute the zero temperature quadrupolar $\chi_Q (\tau) \sim \tau^{-\gamma_Q}$ and octupolar $\chi_O (\tau)\sim \tau^{-\gamma_O}$ susceptibilities with the exponents $\gamma_{Q}$ and $\gamma_{O}$ (see Eq.~\eqref{eq:Qsuscep} and \eqref{eq:Osuscep}), where $\tau$ is the imaginary time. It is shown that the octupolar susceptibility has different scaling behavior around the non-Fermi liquids and critical fixed points at second order in $\epsilon$, which are summarized in Table~\ref{tab:fixed_point_suscep}.
Finally, we propose how the multipolar susceptibilities can be measured experimentally by the use of ultrasound measurements in the presence of a magnetic field. 
The temperature/frequency scaling of the multipolar susceptibility is given by $\chi'_{i}(|\omega/T|\ll1)\sim T^{\gamma_{i}-1}$ and $\chi'_{i}(|\omega/T|\gg1)\sim \omega^{\gamma_{i}-1}$, respectively, where $i=Q,O$ stand for quadrupolar and octupolar, respectively.
The elastic constants are directly related to the multipolar susceptibilties, such as $\Delta(C_{11}-C_{12})\propto \chi'_{Q}$ and $\Delta C_{44}\propto h^{2}\chi'_{O}$ where $\Delta C_{ij}$ is the variation of the elastic constant $C_{ij}$ and $h$ is the magnetic field.
That is, we can obtain the multipolar susceptibility exponent by measuring the temperature dependence of the elastic constants in the presence of the magnetic field using ultrasonic measurements. Our result for the quantum critical behaviors may be realized in experiments on \ce{Pr2(Ti,V)2Al2} in the high pressure regime at low temperature.

The remainder of the paper is organized as follows. In Section~\ref{sec:model}, we describe modelling the multipolar Kondo lattice in terms of a multipolar Bose-Fermi Kondo model. In Section~\ref{sec:rganalysis}, we perform a renormalization group analysis of our multipolar Bose-Fermi Kondo model to identify the phases and phase transitions in the model. In Section~\ref{sec:observables}, we comment on how these phases can be distinguished experimentally, and in Section~\ref{sec:conclusions} we discuss the implications and possible extensions of our work.

\section{Construction of Models}\label{sec:model}

We start by describing the microscopic origin of the multipolar moment, and then describe how to construct conduction orbitals. Then, we couple the multipolar impurity to the conduction orbitalsm which constitutes the (Fermi) Kondo coupling. The Bose-Kondo coupling can then be derived from the Fermi-Kondo coupling by calculating the RKKY coupling of the parent Kondo lattice, and then replacing these other moments with a dynamical bosonic bath field. Since the setting of interest is the prasedoymium cubic compounds \ce{Pr(Ti,V)2Al20}, we need to consider the local symmetry of the \ce{Pr^{3+}} moment. Here, a \ce{Pr^{3+}} ion rests at the centre of a Frank-Kasper cage, which is composed of \ce{(Ti,V)} and \ce{Al}. Despite the complicated nature of the cage, its point group symmetry is simply the tetrahedral group $T_d$. This means that we can classify the wave function of an electron hopping on the cage according to the irreducible representations of $T_d$. We allow the most general interactions according to the local $T_d$ symmetry and time-reversal; the details of the symmetry group are listed in Appendix \ref{app:symmetries}.

\subsection{Multipolar moments}
Generally speaking, on the site of a local moment, the wave functions of a particular ionic configuration are constrained to an effective ground state by Hund's rules. These ground states are then split by the local crystalline electric field. The consequence of these restrictions is the formation of localized anisotropic charge and magnetization densities, leading to multipolar moments. In the case of a rare-earth Pr$^{3+}$ ion subjected to a tetrahedral ($T_d$) crystal field, the spin-orbit coupled $J=4$ multiplet of the 4$f^2$ electrons is split to give rise to a low-lying and energetically well-isolated $\Gamma_{3}$ non-Kramers doublet \cite{Onimaru2016b}; the doublet states are listed in Appendix \ref{app:non_kramers}. This $\Gamma_{3}$ doublet supports both time-reversal even quadrupolar moments $\{\mathcal{O}_{22} = \frac{\sqrt{3}}{2}(J_x^2 - J_y^2)$, $\mathcal{O}_{20} = \frac{1}{2} (3J_z^2 - \bm{J}^2)\}$ as well as a time-reversal odd octupolar moment $\{\mathcal{T}_{xyz} = \frac{\sqrt{15}}{6}\overline{J_xJ_yJ_z}\}$; we use the Stevens operators to describe the multipolar moments and the overline indicates a full symmetrization. These moments can be compactly represented by the pseudospin-1/2 operator $\mathbf{S}$, the components of which are given by \cite{Patri2020d,Schultz2021b}
\begin{equation}
S^x = -\frac{1}{4}\mathcal{O}_{22},\quad S^y = -\frac{1}{4}\mathcal{O}_{20},\quad S^z = \frac{1}{3\sqrt{5}} \mathcal{T}_{xyz},
\end{equation}
and satisfy a canonically normalized $\mathfrak{su}(2)$ algebra $[S^i,S^j] = i\epsilon_{ijk}S^k$. Further details of this pseudospin-1/2 object are described in Appendix \ref{app:non_kramers}. Note that, although the multipolar moments are written in terms of pseudospin-1/2 operators, their transformations under rotations in $T_d$ and time reversal reflect the underlying multipolar attributes. 

\subsection{Fermi-Kondo couplings}
Wave functions of an electron hopping on a Frank-Kasper cage can be thought of as molecular orbitals centred at the Pr ion. It is these molecular orbitals which we couple to the local multipolar moments described in the previous section. Since these wave functions are classifiable according to irreps of $T_d$, we pick basis functions for the $T_2$ representation. Two options are the $p$-orbital basis functions $x,y,z$ (alternatively, or the $T_{2g}$ orbitals $\{xy,yz,zx\}$ yield an identical model and results). We therefore consider three bands, assumed to be degenerate, constructed from these local orbitals; see Eq. \eqref{eq:HF0}. The most general Kondo Hamiltonians coupling of these conduction bands with the local multipolar moments respecting the local $T_d$ symmetry and time-reversal are enumerated in Eqs. \eqref{eq:HQ1}-\eqref{eq:HO} \cite{Patri2020d,Schultz2021b}:

\begin{align}
H_{0}^F ={}&\sum_{\mathbf{k},\alpha,a}E_{\mathbf{k}}c_{\mathbf{k}a\alpha}^{\dagger}c_{\mathbf{k}a\alpha}, \label{eq:HF0} \\
H_{Q1}={}&K_{Q1}c^\dagger_{0a\alpha}\left(\sigma^0_{\alpha\beta} \lambda^3_{ab} S^x - \sigma^0_{\alpha\beta} \lambda^8_{ab} S^y \right) c_{0b\beta}, \label{eq:HQ1}\\
H_{Q2}={}&K_{Q2}c^\dagger_{0a\alpha}\left(2\sigma^z_{\alpha\beta}\lambda^2_{ab} S^y + \sigma^y_{\alpha\beta}\lambda^5_{ab}\left(\sqrt{3}S^x + S^y\right)\right. \label{eq:HQ2} \notag\\
&\left.+ \sigma^x_{\alpha\beta}\lambda^7_{ab}\left(\sqrt{3}S^x - S^y\right) \right) c_{0b\beta}\\
H_{O}={}& K_O c^\dagger_{0 a\alpha}\left(\sigma^x_{\alpha\beta}\lambda^6_{ab} + \sigma^y_{\alpha\beta} \lambda^4_{ab} + \sigma^z_{\alpha\beta}\lambda^1_{ab}\right) S^z c_{0b\beta}. \label{eq:HO}
\end{align}

\noindent The subscript 0 on the conduction electron operators indicates that this interaction occurs only on the impurity site, which is taken to be the origin. The Latin indices sum over orbitals $a,b = x,y,z$, and the Greek indices sum over spins $\alpha,\beta=\uparrow,\downarrow$. $\sigma^i$ are the standard Pauli matrices, and $\lambda^j$ are the $3\times 3$ Gell-Mann matrices, listed in Appendix \ref{app:gell_mann}.  For the  conduction electrons, we assume 
a constant density of states near the Fermi surface, $\sum_{\mathbf{k}}\delta(\omega-E_{\mathbf{k}})=N_{0}$ between $-D<\omega<D$. 

The pseudospin $\mathbf{S}$ represents the multipolar moments, with $S^{x,y}$ and $S^{z}$ standing for the quadrupolar and octupolar moments respectively. In order to perform the many-body perturbation theory later in this work, we rewrite the local moment $\mathbf{S}$ in terms of Abrikosov pseudofermions:
\begin{equation}
\mathbf{S}=\sum_{\alpha\beta}f_{\alpha}^{\dagger}\frac{\vec{\sigma}_{\alpha\beta}}{2}f_{\beta}
\end{equation} 
where we constrain the occupation of the impurity to be $\sum_{\alpha}f_{\alpha}^{\dagger}f_{\alpha}=1$. In order to impose this physical constraint, we introduce a chemical potential for the pseudofermion by adding $\lambda\sum_{\sigma}f^\dagger_\sigma f_\sigma$ to the Hamiltonian, and take the limit $\lambda\to\infty$ at the end of the calculation \cite{Zhu2002, Abrikosov1965}.

\subsection{Bose-Kondo couplings}
In the full Kondo lattice, the local Kondo Hamiltonian of Eqs. \eqref{eq:HQ1}-\eqref{eq:HO} appears at each lattice site. Through this Kondo interaction, an effective interaction between local moments is generated, known as the RKKY interaction \cite{Ruderman1954a, Kasuya1956, Yosida1957}. In the Bose-Fermi Kondo model, this RKKY interaction is represented by the coupling of the local moment to a bosonic bath. The procedure to generate the most general symmetry allowed RKKY-type interaction is described in Appendix \ref{app:rkky}. The resulting kinetic term for bosons and the Bose-Kondo coupling are given in Eqs. \eqref{eq:HB0},\eqref{eq:bose_kondo} respectively,
\begin{align}
H_{0}^B ={}&\sum_{\mathbf{k}}\left[ \Omega_{Q\mathbf{k}}(\phi_{\mathbf{k}}^{x\dagger}\phi^x_{\mathbf{k}}+\phi_{\mathbf{k}}^{y\dagger}\phi^y_{\mathbf{k}}) +\Omega_{O\mathbf{k}}\phi_{\mathbf{k}}^{z\dagger}\phi^z_{\mathbf{k}} \right], \label{eq:HB0} \\
H_{g}={}&g_{Q}(S^{x}\phi^{x}_{0}+S^{y}\phi^{y}_{0})+g_{O}S^{z}\phi^{z}_{0}. \label{eq:bose_kondo}
\end{align}
Here, $\Omega_{Q\mathbf{k}}$ and $\Omega_{O\mathbf{k}}$ are the dispersions of the bosonic baths coupled to the quadrupole and octupole moments, respectively. To set up the controlled RG calculation, we introduce an $\epsilon$ expansion with dimensional regularization in the density of states of the bosonic bath,
\begin{equation}
\sum_{\mathbf{k}}[\delta(\omega-\Omega_{i,\mathbf{k}})-\delta(\omega+\Omega_{i,\mathbf{k}})] = \frac{N_i^2}{2} |\omega|^{1-\epsilon_i}\text{sgn}(\omega).
\end{equation}
To consider the most general situation, we introduce $\epsilon_{Q}$ and $\epsilon_{O}$ for the quadrupolar and octupolar bosonic baths because the density of states power law of the quadrupolar and octupolar bosonic baths are generically different. The multipolar moments localized at $\mathbf{r}=0$ couple to the bosonic bath fields $\vec{\phi}_{0}=\sum_{\mathbf{k}}(\vec{\phi}_{\mathbf{k}}+\vec{\phi}_{-\mathbf{k}}^{\dagger})$. 

In summary, the total multipolar Bose-Fermi Kondo Hamiltonian $H$ is
\begin{align}
    H={}&H_{0}^{F}+H_{Q1}+H_{Q2}+H_{O}+H_{0}^{B}+H_{g}.
\end{align}

\section{Renormalization Group Analysis}\label{sec:rganalysis}

\subsection{$\epsilon$-Expansion and Dimensional Regularization}
We perform the renormalization group analysis by using dimensional regularization with minimal subtraction \cite{Zhu2002}. The bosonic bath already has an $\epsilon$ factor modifying its density of states which can be used in the minimal subtraction procedure, but the conduction electron bath does not. We therefore introduce $\epsilon'$ for the conduction electron density of states to enable the minimal subtraction of poles:
\begin{align}
    \sum_{\mathbf{k}}\delta(\omega-E_{\mathbf{k}})=N_{0}|\omega|^{-\epsilon'}.
\end{align}
Note that $\epsilon'$ will set to zero at the end of the calculation. Consequently, we define a renormalized field $f$ and dimensionless coupling constants $g_{i}$ and $K_{i}$,
\begin{align}
    f^{B}={}&Z_{f}^{1/2}f,\\
    g^{B}_{i}={}&g_{i} Z_{f}^{-1}Z_{g_{i}}\mu^{\epsilon_{i}/2},\\
    K^{B}_{j}={}&K_{j} Z_{f}^{-1}Z_{K_{j}}\mu^{\epsilon'},
\end{align}
where $\mu$ is the renormalization energy scale, and $Z_{f}$, $Z_{g_i}$, and $Z_{K_j}$ are the renormalization constants for the pseudofermion $f$, bosonic couplings $g_{i}$ (here $i=Q,O$), and fermionic couplings $K_j$ (here $j=Q_{1},Q_{2},O$). The superscript $B$ stands for the bare value which does not evolve under the RG flow. In addition, we absorb the density of states $N_{i}$ into the dimensionless couplings as $N_{0}K_{j}\rightarrow K_{j}$ and $N_{i}g_{i}\rightarrow g_{i}$, respectively, in the following section. The details of the RG analysis and corresponding Feynman diagrams are enumerated in Appendix \ref{app:rg}.
Note that we ignore the self-energies of the conduction electrons and bosonic baths because they vanish in the thermodynamic limit \cite{Si2003}.

\subsection{Analysis of the Fermionic Kondo Model} \label{sec:fkondo}
The beta functions with the multipolar moment couplings up to cubic order in $K_{i}$ are given by \cite{Patri2020e, Schultz2021b, Patri2020d}
\begin{align}
\frac{dK_{Q1}}{d\ln\mu}={}&6K_{Q2}K_{O}+2K_{Q1}(K_{Q1}^{2}+6K_{Q2}^{2}+3K_{O}^{2}),\\
\frac{dK_{Q2}}{d\ln\mu}={}&K_{O}(K_{Q1}-\sqrt{3}K_{Q2}) \\&+2K_{Q2}(K_{Q1}^{2}+6K_{Q2}^{2}+3K_{O}^{2}), \nonumber \\
\frac{dK_{O}}{d\ln\mu}={}&2K_{Q2}(2K_{Q1}-\sqrt{3}K_{Q2}) \\
&+4K_{O}(K_{Q1}^{2}+6K_{Q2}^{2}). \nonumber
\end{align}
This RG flow has two distinct stable fixed points. 
The two types of stable fixed points are the multipolar fixed points, $F_{1\pm}=(K_{Q1}^{*},K_{Q2}^{*},K_{O}^{*})=(\pm\frac{1}{2\sqrt{6}},\pm\frac{1}{12\sqrt{2}},-\frac{1}{4\sqrt{3}})$, and two-channel Kondo fixed points, $F_{2\pm}=(\pm\frac{1}{2\sqrt{3}},\mp\frac{1}{6},\frac{1}{2\sqrt{3}})$. 
The stable fixed points have perturbative scaling dimensions $\Delta=1/4$ and $\Delta=1$, respectively, which are the slope of the beta function at the respective fixed points; both fixed points are non-Fermi liquid phases. $\Delta$ is also related to the scaling dimension $(1+\Delta)$ of the leading irrelevant operator at the fixed point. The physical observables such as resistivity $\rho$ and heat capacity $C_V$ at the fixed points are obtained by using the scaling dimension $\Delta$; at low temperatures we have $\rho\sim T^\Delta$ and $C_V \sim T^{2\Delta}$. Note that the exact scaling dimensions of $F_{1\pm}$ and $F_{2\pm}$ from the CFT are $1/5$ and $1/2$, respectively \cite{Affleck1993b,Patri2020e, Patri2020d}.

\subsection{Analysis of the Bosonic Kondo Model}\label{sec:bkondo}
The beta functions for the coupling of the local moment to the bosonic bath up to $g_{i}^{5}$ order are given by
\begin{align}
\frac{d\lambda_{Q}}{d\ln\mu}={}&-\lambda_{Q}\left(\epsilon_{Q}-\lambda_{Q}-\lambda_{O}+\lambda_{Q}^{2}+\lambda_{Q}\lambda_{O}\right),
\label{eq:lambdaq0}\\
\frac{d\lambda_{O}}{d\ln\mu}={}&-\lambda_{O}\left(\epsilon_{O}-2\lambda_{Q}+2\lambda_{Q}\lambda_{O}\right),\label{eq:lambdao0}
\end{align}
where $\lambda_{Q,O}=g_{Q,O}^{2}$. Eqs.~\eqref{eq:lambdaq0} and \eqref{eq:lambdao0} have two stable fixed points, a quadrupolar ordered fixed point, $B_{Q}=(\lambda_{Q}^{*},\lambda_{O}^{*})=(\epsilon_{Q}+\epsilon_{Q}^{2},0)$, and an octupolar ordered fixed point, $B_{O}=(0,\infty)$. The quadrupolar and octupolar fixed points can be identified with an $XY$ fixed point and Ising fixed point in the ordinary Fermi-Bose Kondo model \cite{Zhu2002,Zarand2002}. The octupolar fixed point is, strictly speaking, outside of the regime of our perturbative calculation. 
The beta functions also have another fixed point, $C_{QO}=(\frac{\epsilon_{Q}}{2}+\frac{\epsilon_{O}(2\epsilon_{Q}-\epsilon_{O})}{4},\frac{(2\epsilon_{Q}-\epsilon_{O})}{2}+\frac{\epsilon_{O}^{2}}{4})$, which is a critical point between the quadrupolar and octupolar fixed points, and corresponds to the XXZ fixed point in the ordinary Fermi-Bose Kondo model \cite{Zhu2002,Zarand2002}. All the fixed point values are calculated up to $\epsilon_{i}^{2}$ order. In the limit $\epsilon_{Q}=\epsilon_{O}=\epsilon$, $C_{QO}=(\frac{\epsilon}{2}+\frac{\epsilon^{2}}{4},\frac{\epsilon}{2}+\frac{\epsilon^{2}}{4})$ becomes isotropic  \cite{Zhu2002, Zarand2002}. Note that $B_{Q}$ exists in the range $0<\epsilon_{Q}<1/4$, and $C_{QO}$ exists when $27\epsilon_{O}^{2}-2(2+\epsilon_{O}-2\epsilon_{Q})^{3}<0$ and $\epsilon_{Q}\gtrsim\sqrt{3.516+2\epsilon_{O}-\epsilon_{O}^{2}}-1.875$ with $0<\epsilon_{O}<1$. For the isotropic limit $\epsilon = \epsilon_{Q} = \epsilon_{O}$, the range is $0<\epsilon<1/2$.

\subsection{Analysis of the Bose-Fermi Kondo Model}\label{sec:bfkondo}
\begin{figure}[h]
\includegraphics[scale=0.85]{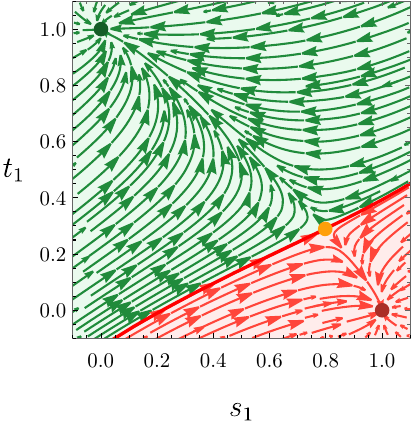}
\caption{The RG flow diagram between the multipolar Kondo fixed point $F_{1+}$ (green dot) and quadrupolar fixed point $B_{Q}$ (red dot) when $\epsilon=0.1$. $F_{1+}$ and $B_{Q}$ are located at $(s_{1},t_{1})=(0,1)$ and $(s_{1},t_{1})=(1,0)$, respectively. Between the two stable fixed points, there is a critical point $C_{1+}^{Q}=(0.798, 0.289)$ (orange dot), and the red line denotes the separatrix between these two phases. Here, $s_{1}=1-0.104K_{Q1}-0.180K_{Q2}+6.708K_{O}$, $t_{1}=3.796K_{Q1}+6.574K_{Q2}+1.124K_{O}$, and $\lambda_{O}=0$, with the constraint $\lambda_{Q}=0.113-0.571K_{Q1}-0.989K_{Q2}-0.430K_{O}$.
}\label{fig:phase1}
\end{figure}

\begin{figure}[h]
\includegraphics[scale=0.85]{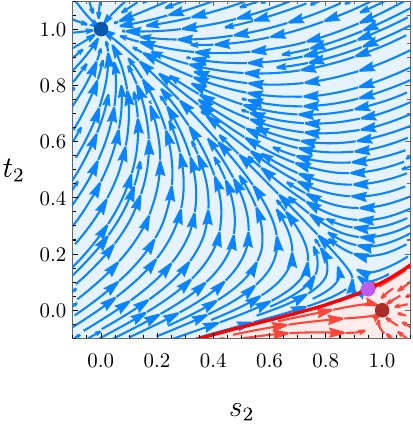}
\caption{The RG flow diagram between the 2-channel Kondo fixed point $F_{2+}$ (blue dot, $(s_{2},t_{2})=(0,1)$) and quadrupolar fixed point $B_{Q}$ (red dot, $(s_{2},t_{2})=(1,0)$) when $\epsilon=0.1$. Between the two fixed points, there is a critical point $C_{2+}^{Q}=(0.950, 0.075)$ (purple dot). The red line denotes the separatrix between the two phases. Here, $s_{2}=1-0.003K_{Q1}+0.011K_{Q2}-3.455K_{O}$, $t_{2}=1.219K_{Q1}-4.222K_{Q2}-0.192K_{O}$, and $\lambda_{O}=0$ with the constraint $\lambda_{Q}=0.113-0.218K_{Q1}+0.755K_{Q2}+0.263K_{O}$.}\label{fig:phase2}
\end{figure}

In order to study the destruction of the Kondo effect to magnetic ordering, we consider the full model of coupling the local moment to both the fermionic conduction electron bath and the bosonic bath. In this case, the beta functions are as follows:
\begin{align}
\frac{dK_{Q1}}{d\ln\mu}={}&K_{Q1}\left(\frac{\lambda_{Q}+\lambda_{O}}{2}-\frac{\lambda_{Q}(\lambda_{Q}+\lambda_{O})}{2}\right)\\
&+6K_{Q2}K_{O}+2K_{Q1}(K_{Q1}^{2}+6K_{Q2}^{2}+3K_{O}^{2}), \nonumber \\
\frac{dK_{Q2}}{d\ln\mu}={}&K_{Q2}\left(\frac{\lambda_{Q}+\lambda_{O}}{2}-\frac{\lambda_{Q}(\lambda_{Q}+\lambda_{O})}{2}\right)\\
&+K_{O}(K_{Q1}-\sqrt{3}K_{Q2})\notag\\
&+2K_{Q2}(K_{Q1}^{2}+6K_{Q2}^{2}+3K_{O}^{2}), \nonumber \\
\frac{dK_{O}}{d\ln\mu}={}&K_{O}(\lambda_{Q}-\lambda_{Q}\lambda_{O})\\
&+2K_{Q2}(2K_{Q1}-\sqrt{3}K_{Q2})+4K_{O}(K_{Q1}^{2}+6K_{Q2}^{2}), \nonumber \\
\frac{d\lambda_{Q}}{d\ln\mu}={}&-\lambda_{Q}[\epsilon_{Q}-(\lambda_{Q}+\lambda_{O})+\lambda_{Q}(\lambda_{Q}+\lambda_{O})\label{eq:beta_lambdaQ}\\
&-4(K_{Q1}^{2}+6K_{Q2}^{2}+3K_{O}^2)], \nonumber \\
\frac{d\lambda_{O}}{d\ln\mu}={}&-\lambda_{O}[\epsilon_{O}-2\lambda_{Q}+2\lambda_{Q}\lambda_{O}-8(K_{Q1}^{2}+6K_{Q2}^{2})].\label{eq:beta_lambdaO}
\end{align}
Under this full renormalization group flow, all of the previously found stable fixed points in the fermionic Kondo $F_{1\pm}$, $F_{2\pm}$ and bosonic Kondo $B_{Q}$, and $B_{O}$ cases remain stable. Further, new fixed points emerge, which describe critical points between the phases described in Secs.~\ref{sec:fkondo}-\ref{sec:bkondo}. For the case $\lambda_Q\neq0, \lambda_O = 0$, we find two pairs of critical points. 
The first critical point is given by $C^{Q}_{1\pm}=(K_{Q1}^{*},K_{Q2}^{*},K_{O}^{*},\lambda_{Q}^{*},\lambda_{O}^{*})=(\pm\frac{\epsilon_{Q}}{\sqrt{3}},\pm\frac{\epsilon_{Q}}{6},-\frac{\epsilon_{Q}}{2\sqrt{3}},\epsilon_{Q}-2\epsilon_{Q}^{2},0)$ which is a critical point between $F_{1\pm}$ and $B_{Q}$. The flow diagram corresponding to this transition is given in Fig.~ \ref{fig:phase1}. The second critical point is $C^{Q}_{2\pm}=(\pm(\frac{\epsilon_{Q}}{2\sqrt{6}}+\frac{\sqrt{3}\epsilon_{Q}^{2}}{16\sqrt{2}}),\mp(\frac{\epsilon_{Q}}{6\sqrt{2}}+\frac{\epsilon_{Q}^{2}}{16\sqrt{2}}),\frac{\epsilon_{Q}}{4\sqrt{3}},\epsilon_{Q}+\frac{\epsilon_{Q}^{2}}{4},0)$. This is a critical point between $F_{2\pm}$ and $B_{Q}$, and its flow diagram is Fig.~\ref{fig:phase2}.
In the case $\lambda_{Q}=0$, $\lambda_O \neq 0$, we find the critical points, $C_{1,2\pm}^{O}$ , between $F_{1,2\pm}$ and $B_{O}$. However, the fixed point values of $\lambda_{O}$ in  $B_{O}$ and $C_{1,2\pm}^{O}$ are order one numbers, so they are outside of the perturbative regime. Despite this, we believe the the existence of the fixed points to be maintained under specalized nonperturbative methods. For example, the octupolar-type critical points may be accessible via the Coulomb gas representation \cite{Si1996,Zhu2002}.
 Note that $C_{1\pm}^{Q,O}$ and $C_{2\pm}^{Q,O}$ exist for $0<\epsilon_{Q}<1/2$ and $0<\epsilon_{Q}<2-\sqrt{2}$, respectively.

\begin{table*}[!ht]
\centering
\begin{tabular}{|>{$}c<{$}|>{$}c<{$}|c|>{$}c<{$}|>{$}c<{$}|}
\hline
\text{Label}&(K_{Q1}^{*},K_{Q2}^{*},K_{O}^{*},\lambda_{Q}^{*},\lambda_{O}^{*})&Type&\gamma_{Q}&\gamma_{O}\\
\hline\hline
F_{1\pm}&\left(\pm\frac{1}{2\sqrt{6}},\pm\frac{1}{12\sqrt{2}},-\frac{1}{4\sqrt{3}},0,0\right)& Multipolar & 1/2 & 1/2\\
F_{2\pm}&\left(\pm\frac{1}{2\sqrt{3}},\mp\frac{1}{6},\frac{1}{2\sqrt{3}},0,0\right)&2-channel Kondo&2&2\\
B_{Q}&\left(0,0,0,\epsilon_{Q}+\epsilon_{Q}^{2},0\right)&Quadrupolar&\epsilon_{Q}&2\epsilon_{Q}+2\epsilon_{Q}^{2}\\
B_{O}&\left(0,0,0,0,\infty\right)&Octupolar&-&\epsilon_{O}\\
C_{QO}&\left(0,0,0,\frac{\epsilon_{O}}{2}+\frac{\epsilon_{O}(2\epsilon_{Q}-\epsilon_{O})}{4},\frac{(2\epsilon_{Q}-\epsilon_{O})}{2}+\frac{\epsilon_{O}^{2}}{4}\right)& Critical&\epsilon_{Q}&\epsilon_{O}\\
\hline
C_{1\pm}^{Q}&\left(\pm\frac{\epsilon_{Q}}{\sqrt{3}},\pm\frac{\epsilon_{Q}}{6},-\frac{\epsilon_{Q}}{2\sqrt{3}},\epsilon_{Q}-2\epsilon_{Q}^{2},0\right)&Critical&\epsilon_{Q}&2\epsilon_{Q}\\
C_{2\pm}^{Q}&\left(\pm\big(\frac{\epsilon_{Q}}{2\sqrt{6}}+\frac{\sqrt{3}\epsilon_{Q}^{2}}{16\sqrt{2}}\big),\mp\big(\frac{\epsilon_{Q}}{6\sqrt{2}}+\frac{\epsilon_{Q}^{2}}{16\sqrt{2}}\big),\frac{\epsilon_{Q}}{4\sqrt{3}},\epsilon_{Q}+\frac{\epsilon_{Q}^{2}}{4},0\right)&Critical&\epsilon_{Q}&2\epsilon_{Q}+{3\epsilon_{Q}^{2}}/{2}\\
C_{1\pm}^{O}&\left(\pm\frac{\epsilon_{O}^{1/2}}{2\sqrt{3}},\pm\frac{\epsilon_{O}^{1/2}}{12},-\frac{1}{4\sqrt{3}},0,\frac{1}{4}-\frac{\epsilon_{O}}{2}\right)&\text{Critical}&-&\epsilon_{O}\\
C_{2\pm}^{O}&\left(\pm\frac{\epsilon_{O}^{1/2}}{2\sqrt{6}},\mp\frac{\epsilon_{O}^{1/2}}{6\sqrt{2}},\frac{1}{2\sqrt{3}},0,1-\frac{\epsilon_{O}}{2}\right)&\text{Critical}&-&\epsilon_{O}\\
\hline
\end{tabular}
\caption{Table of the fixed points and their multipolar susceptibility exponents.
$F_{1\pm}$ and $F_{2\pm}$ are the multipolar and 2-channel Kondo fixed points, respectively, and $B_{Q}$ and $B_{O}$ are the quadrupolar fixed point and octupolar fixed line, respectively (all four of these are stable).
$C_{1,2\pm}^{Q}$ is the critical point between $F_{1,2\pm}$ and $B_{Q}$, and $C_{1,2\pm}^{O}$ is the critical point between $F_{1,2\pm}$ and $B_{O}$.
$C_{QO}$ is the critical point between $B_{Q}$ and $B_{O}$. 
$\gamma_{Q}$ and $\gamma_{O}$ stand for the quadrupolar and octupolar susceptibility exponents defined in Eq.~\eqref{eq:Qsuscep} and \eqref{eq:Osuscep}, respectively. 
The schematic diagram for their quantum phase transitions is presented in Fig.~\ref{fig:destruction}.
Note that $B_{O}$ and $C_{1,2\pm}^{O}$ are outside of the perturbative regime, so their $\gamma_{Q}$ values are omitted.
Note that in this table, the multipolar susceptibility exponents of $F_{(1,2)\pm}$ are the perturbative values. The exact values from conformal field theory are $2/5$ and $1$, respectively \cite{Affleck1993b,Patri2020e}.
}
\label{tab:fixed_point_suscep}
\end{table*}

\section{Physical observables}\label{sec:observables}

\subsection{Multipolar Susceptibility and Exponents}

In order to compare our results with experiment, we consider the local multipolar moment susceptibility exponent. The local quadrupolar and octupolar moment susceptibilities, $\chi_{Q}$ and $\chi_{O}$, are defined as
\begin{align}
\chi_{Q}(\tau)={}&\braket{T_{\tau}S^{x,y}(\tau)S^{x,y}(0)}\propto\left(\frac{\tau_{0}}{|\tau|}\right)^{\gamma_{Q}},\label{eq:Qsuscep}\\
\chi_{O}(\tau)={}&\braket{T_{\tau}S^{z}(\tau)S^{z}(0)}\propto\left(\frac{\tau_{0}}{|\tau|}\right)^{\gamma_{O}},\label{eq:Osuscep}
\end{align}
where $\gamma_{i}$ ($i=Q,O$) is the multipolar susceptibility exponent and $\tau\gg\tau_{0}$ with the cutoff $\tau_{0}=1/\Lambda$. We emphasize that the multipolar susceptibility exponents describe how the susceptibility scales as imaginary time evolves, but do not directly yield the temperature scaling.
When the fixed point value of $\lambda_{i}$ ($i=Q,O$) is non-zero, the corresponding susceptibility exponent is given by \cite{Zhu2002,Zarand2002}
\begin{align}
\gamma_{i}=\epsilon_{i}+\left[\frac{1}{\lambda_{i}}\frac{d\lambda_{i}}{d\ln \mu}\right]_{\text{f.p.}},
\end{align}
where f.p. stands for value at the fixed point.
By definition, $\frac{d\lambda_{i}}{d\ln\mu}=0$ at the fixed point, so $\gamma_{i}=\epsilon_{i}$, which is exact to all orders of $\epsilon$ \cite{Zhu2002,Zarand2002}. Since our critical points $C_{1,2\pm}^{Q}$
all have a non-zero fixed point value for $\lambda_{Q}$, the quadrupolar susceptibility exponent of the critical points $C_{1\pm}^{Q}$ and $C_{2\pm}^{Q}$ are both given by $\gamma_{Q}=\epsilon_{Q}$. Thus we cannot distinguish between the two critical points via $\gamma_{Q}$. Let us consider instead the case of the octupolar susceptibility exponent.

When the fixed point value of $\lambda_{i}$ is zero, as is the case for $\lambda_O$ at both $C_{1,2\pm}^{Q}$, then the corresponding susceptibility exponent is given by \cite{Zarand2002}
\begin{align}
\gamma_{i}={}&\epsilon_{i}+\left[\lim_{\lambda_{i}\rightarrow0}\frac{1}{\lambda_{i}}\frac{d\lambda_{i}}{d\ln \mu}\right]_{\text{f.p.}}\notag\\
={}&\epsilon_{i}+\left[\frac{\partial}{\partial\lambda_{i}}\frac{d\lambda_{i}}{d\ln \mu}\right]_{\text{f.p.}}.\label{eq:suscep_zero}
\end{align}
In contrast to the previous case, Eq.~\eqref{eq:suscep_zero} includes higher order corrections in $\epsilon_{Q,O}$. Since our calculation applies to order $\epsilon^2$, we can use $\epsilon^2$ corrections to the octupolar susceptibility to distinguish between different fixed points. The resulting susceptibilities for the two critical points are $\gamma_{O}=2\epsilon_Q$ and $2\epsilon_Q+3\epsilon_Q^{2}/2$, for $C^{Q}_{1\pm}$ and $C^{Q}_{2\pm}$, respectively. The results for the susceptibility exponents at different fixed points are summarized at Table~\ref{tab:fixed_point_suscep}. The full expression of the multipolar susceptibility exponent $\gamma_{i}$ for $\lambda_{i}^{*}=0$ is presented in Appendix \ref{app:multipolar_susc}. An additional point is that we may also distinguish between these critical points and the non-Fermi liquid phases $F_{1\pm}$ and $F_{2\pm}$ using this octupolar susceptibility. This is useful because we can then distinguish non-Fermi liquid behavior due to a quantum critical regime from non-Fermi liquid behavior in a phase ($F_{1\pm}$ or $F_{2\pm}$ in our case). We note that the multipolar susceptibility exponents of the fermionic fixed points $F_{(1,2)\pm}$ are perturbative. The exact exponents can be obtained via conformal field theory and are $2/5$ and $1$ for $F_{1\pm}$ and $F_{2\pm}$, respectively \cite{Affleck1993b,Patri2020e}.

\subsection{Finite Temperature Scaling and Elastic Constants}

The results in the previous section only apply at zero temperature, and do not directly correspond to a measurable quantity. In order to obtain the temperature dependence of the susceptibility, we assume that we have conformal invariance at the critical point, and that the multipolar moment is a primary operator with conformal dimension $\gamma_{i}/2$ \cite{Zarand2002,Aronson1997}. The results for the real part $\chi'$ and imaginary part $\chi''$ are given by Eqs.~\eqref{eq:real_susc}-\eqref{eq:imag_susc}; see Appendix \ref{app:temp_suscep} for details.
\begin{align}
    \chi_{i}'(\omega,T)\propto{}&\begin{cases}
    T^{\gamma_{i}-1}\left(1+C_{\text{Re}1}\left(\frac{\omega}{T}\right)^{2}\right),&|\omega|\ll T,\\
    \omega^{\gamma_{i}-1},&|\omega|\gg T,\\
    \end{cases} \label{eq:real_susc}\\
    \chi_{i}''(\omega,T)\propto{}&\begin{cases}
    T^{\gamma_{i}-1}\left(\frac{\omega}{T}\right),&|\omega|\ll T,\\
    \omega^{\gamma_{i}-1},&|\omega|\gg T,\\
    \end{cases} \label{eq:imag_susc}
\end{align}
where $C_{\text{Re}1}$ is defined in Appendix~\ref{app:temp_suscep}.
Note that the real part of the temperature dependence of the multipolar susceptibility for $F_{1\pm}$ and $F_{2\pm}$ in the dc limit ($\omega=0$) is consistent with the temperature scaling, $\chi\sim T^{2\Delta-1}$, which is the conformal field theory result upon setting $\gamma_{i} = 2\Delta$ \cite{Affleck1993b}. The exact temperature dependencies for $F_{1\pm}$ and $F_{2\pm}$ from CFT are given by $\chi\sim T^{-3/5}$ and $\chi\sim\log T$, respectively \cite{Patri2020e}.
We expect that the multipolar susceptibility exponent in Eq.~\eqref{eq:real_susc} can be observed by measuring the temperature dependence of elastic constants \cite{Yanagisawa2018,Sorensen2021}. 
The elastic free energy including the symmetry-allowed coupling between the multipolar moments and strains is given by \cite{Sorensen2021, Patri2019d}
\begin{align}
\mathcal{F}={}&\frac{C_{11}^{0}-C_{12}^{0}}{2}(\epsilon_{\mu}^{2}+\epsilon_{\nu}^{2})+\frac{C_{44}^{0}}{2}(\epsilon_{xy}^{2}+\epsilon_{yz}^{2}+\epsilon_{zx}^{2})\notag\\
&-s_{Q}[\epsilon_{\mu}\mathcal{O}_{22}+\epsilon_{\nu}\mathcal{O}_{20}]\notag\\
&-s_{O}\mathcal{T}_{xyz}[h_{x}\epsilon_{yz}+h_{y}\epsilon_{zx}+h_{z}\epsilon_{xy}],
\end{align}
where $\epsilon_{ij}$ is the strain tensor, $\epsilon_{\mu}\equiv(\epsilon_{xx}-\epsilon_{yy})$ and $\epsilon_{\nu}\equiv(2\epsilon_{zz}-\epsilon_{xx}-\epsilon_{yy})/\sqrt{3}$, $C_{11}$, $C_{12}$, $C_{44}$ are the elastic constants which are coefficients of $\epsilon_{ii}^{2}$, $\epsilon_{ii}\epsilon_{jj}$, $\epsilon_{ij}^{2}$ ($i\neq j$) for the deformation free energy in the cubic lattice, respectively; $h_{i}$ is the magnetic field ($i=x,y,z$); the superscript $0$ stands for the bare value of the elastic constants; $s_Q$ and $s_O$ are the couplings between the multipolar moments and lattice strain tensors. From second-order perturbation theory, we can get the following corrections to the bare elastic constants,
\begin{align}
(C_{11}-C_{12})={}&(C_{11}^{0}-C_{12}^{0})-(s_{Q}^{2})\chi'_{Q},\\
C_{44}={}&C_{44}^{0}-(s_{O}^{2}h_{i}^{2})\chi'_{O},
\end{align}
where we apply the magnetic field $h_{i}$ along one and only one of the $x$, $y$, or $z$ axes under the assumption that the cubic symmetry is negligbly affected. 
The octupolar susceptibility is therefore only detectable when measured in the presence of both strain and magnetic field simultaneously. 
As a result, the multipolar susceptibility can be observed by measuring the temperature dependence of the elastic constants $(C_{11}-C_{12})$ and $C_{44}$ via ultrasonic measurements.

\section{Conclusions}\label{sec:conclusions}
Inspired by experiments on \ce{Pr(Ti,V)2Al20}, we have studied an appropriate multipolar Bose-Fermi Kondo model, where the non-Kramers local moments carrying quadrupolar and octupolar moments are coupled to $p$-orbital electrons. 
By using an RG analysis on our model, we find not only two non-Fermi liquid phases and a quadrupolar ordered phase, but also two quantum critical points between the non-Fermi liquid phases and quadrupolar ordered phase.
To distinguish between each of these non-Fermi liquid phases and quantum critical points, we compute the multipolar susceptibility exponents at zero temperature and show that the octupolar susceptibility exponent is different at second order in $\epsilon$ at all of these fixed points.
Furthermore, we obtain the temperature scaling behavior of the multipolar susceptibility, and explain how the quadrupolar and octupolar susceptibilities are related to the elastic constants $(C_{11}-C_{12})$ and $C_{44}$, respectively. We propose that measurement of the temperature dependence of the elastic constants $(C_{11}-C_{12})$ and $C_{44}$ using an ultrasonic measurement in the presence of a magnetic field can be used to distinguish the non-Fermi liquid phases and quantum critical points experimentally. Our results may be experimentally relevant for \ce{Pr2(Ti,V)2Al2} in the high-pressure regime at low temperature.

Possible directions for future work could include applying the work to a variety of other heavy fermion systems. For example, several Yb and Ce compounds exhibit local moments with very high degeneracies, which enables the formation of a large number of multipolar moments \cite{Rosenberg2019c, Martelli2019, Tsuruta2000, Nakamura2019, Falkowski2014, Schultz2021c, Liu2021, Shiina1998, Shiina1997}. Another direction could be to verify our results from the (extended) dynamical mean field theory perspective. One subtlety is that the $\epsilon_Q,\epsilon_O$ parameters in the bosonic bath density of states should be determined self-consistently, which may be possible in a full dynamical mean field treatment \cite{Smith2000}.

More generally, our results are indicative of the large variety of multipolar ordered phases and exotic electronic states found in rare-earth metallic systems. The root of the multipolar moments, unusual Kondo couplings, and anisotropic RKKY interactions is the strong spin-orbit coupling and crystal electric field effects, which, as we have shown, can lead to a myriad of quantum critical behaviors.
This suggests there may be new classes of quantum critical points relating Kondo destruction, multipolar ordering, and non-Fermi liquids in multipolar Kondo lattice systems, and that they are experimentally accessible. This opens new doors for exploring the landscape of multipolar quantum matter. 

\begin{acknowledgments}
This work was supported by the NSERC of Canada and the
Center for Quantum Materials at the University of Toronto. Y.B.K. is also supported by the Simons Fellowship from the Simons Foundation and the Guggenheim Fellowship from the John Simon Guggenheim Memorial Foundation. D.S. is supported by the Ontario Graduate Scholarship.
\end{acknowledgments}

\appendix
\section{Multipolar moments from Non-Kramers doublet \label{app:non_kramers} }
In a vacuum, a \ce{Pr^{3+}} ion forms a spin $J=4$ system by Hund's rules. In the presence of a tetrahedral crystal field, these 9 degenerate states are split, and the resulting ground state in the \ce{Pr(V,Ti)_2Al_{20}} compounds is a non-Kramers $\Gamma_3$ doublet spanned by the following two states \cite{Onimaru2016b,Patri2020d,Schultz2021b}:
\begin{align}
\ket{\Gamma^{(1)}_3} ={}& \frac{1}{2}\sqrt{\frac{7}{6}}\ket{4} - \frac{1}{2}\sqrt{\frac{5}{3}}\ket{0} + \frac{1}{2}\sqrt{\frac{7}{6}}\ket{-4}, \\
\ket{\Gamma^{(2)}_3} ={}& \frac{1}{\sqrt{2}}\ket{2} + \frac{1}{\sqrt{2}} \ket{-2}.
\end{align}
To determine which multipolar moments are supported by these wave functions, we can compute the matrix elements of Stevens operators in the doublet $\{\ket{\Gamma^{(1)}_3}, \ket{\Gamma^{(2)}_3}\}$. In this doublet, we find that, defining a different basis
\begin{align}
\ket{\uparrow} ={}& \frac{1}{\sqrt{2}}\left(\ket{\Gamma^{(1)}_3}+i \ket{\Gamma^{(2)}_3}\right), \label{eq:spin_up} \\
\ket{\downarrow} ={}& \frac{1}{\sqrt{2}}\left(i\ket{\Gamma^{(1)}_3}+ \ket{\Gamma^{(2)}_3}\right),\label{eq:spin_down}
\end{align}
we find that 
\begin{align}
\bra{\alpha}\left(-\frac{1}{4}\mathcal{O}_{22} \right)\ket{\beta} ={}& \frac{1}{2}\sigma^x_{\alpha\beta},\\
\bra{\alpha}\left(-\frac{1}{4}\mathcal{O}_{20} \right)\ket{\beta} ={}& \frac{1}{2}\sigma^y_{\alpha\beta},\\
\bra{\alpha}\left(\frac{1}{3\sqrt{5}}\mathcal{T}_{xyz} \right)\ket{\beta} ={}& \frac{1}{2}\sigma^z_{\alpha\beta},
\end{align}
where $\alpha,\beta$ take the values $\uparrow,\downarrow$ (where these $\uparrow,\downarrow$ are the ones listed in Eqs.~\eqref{eq:spin_up}-\eqref{eq:spin_down}), and $\sigma^i$ are the standard Pauli matrices. We emphasize that these $\sigma^i$ matrices and indices $\alpha,\beta$ relate to matrix elements of operators in local moment states, and have nothing to do with the Pauli matrices and $\alpha,\beta$ indices for the conduction electrons in Eqs.~\eqref{eq:HQ1}-\eqref{eq:HO}.

\section{\label{app:symmetries} Action of Tetrahedral Group}
In order to test which terms in the Hamiltonian are allowed, we need to know how candidate terms transform under action of the tetrahedral group $T_d$, and under time-reversal $\mathcal{T}$. The most economical way to check all transformations is pick two generators of $T_d$, which are $\mathcal{C}_{31}$ and $\mathcal{S}_{4z}$. $\mathcal{C}_{31}$ is the rotation by $2\pi/3$ about the $(1,1,1)$ axis, and $S_{4z}$ is a rotation by $\pi/2$ about the $z$-axis followed by a mirror reflection across the $xy$ plane. Both of these transformations leave a tetrahedron invariant. Checking all possible Kondo terms respecting the symmetry yields Eqs.~\eqref{eq:HQ1}-\eqref{eq:HO}. The table of all symmetry transformation is given by Table \ref{tab:symmetries}.

\begin{table}[ht]
\centering
\begin{tabular}{c|c|c|c}
Object & $\mathcal{S}_{4z}$ & $\mathcal{C}_{31}$ & $\mathcal{T}$ \\ \hline\hline
$x$ & $-y$ & $y$ & $x$ \\
$y$ & $x$ & $z$ & $y$ \\
$z$ & $-z$ & $x$ & $z$ \\ \hline
$\sigma^0$ & $\sigma^0$ & $\sigma^0$ & $\sigma^0$ \\
$\sigma^x$ & $\sigma^y$ & $\sigma^y$ & $-\sigma^x$ \\
$\sigma^y$ & $-\sigma^x$ & $\sigma^z$ & $-\sigma^y$ \\
$\sigma^z$ & $\sigma^z$ & $\sigma^x$ & $-\sigma^z$ \\ \hline
$S^x$ & $-S^x$ & $-\frac{1}{2}S^x - \frac{\sqrt{3}}{2}S^y$ & $S^x$ \\ 
$S^y$ & $S^y$ & $\frac{\sqrt{3}}{2}S^x - \frac{1}{2}S^y$ & $S^y$ \\ 
$S^z$ & $-S^z$ & $S^z$ & $-S^z$ \\ 
\end{tabular}
\caption{Symmetry transformations of various objects under two generators of the tetrahedral group as well as time-reversal $\mathcal{T}$. \label{tab:symmetries}}
\end{table}

\section{$\text{SU}(3)$ Gell-Mann Matrices \label{app:gell_mann}}
In our multipolar Kondo models, we have three orbitals. To account for all possible traceless hermitian matrices which describe possible fermionic bilinears, we use the generators of $\text{SU}(3)$, normalized to $\text{tr}(\lambda^i\lambda^j) = 2\delta_{ij}$. We enumerate these $3\times 3$ Gell-Mann matrices that appear in the Fermi-Kondo Hamiltonians here \cite{Patri2020d,Schultz2021b}:
\begin{align}
\lambda^1 =& \begin{pmatrix} 0 & 1 & 0 \\ 1 & 0 & 0 \\ 0 & 0 & 0 \end{pmatrix}, 
&\lambda^2 =& \begin{pmatrix} 0 & -i & 0 \\ i & 0 & 0 \\ 0 & 0 & 0 \end{pmatrix}, \\
\lambda^3 =& \begin{pmatrix} 1 & 0 & 0 \\ 0 & -1 & 0 \\ 0 & 0 & 0 \end{pmatrix}, 
&\lambda^4 =& \begin{pmatrix} 0 & 0 & 1 \\ 0 & 0 & 0 \\ 1 & 0 & 0 \end{pmatrix}, \\
\lambda^5 =& \begin{pmatrix} 0 & 0 & -i \\ 0 & 0 & 0 \\ i & 0 & 0 \end{pmatrix}, 
&\lambda^6 =& \begin{pmatrix} 0 & 0 & 0 \\ 0 & 0 & 1 \\ 0 & 1 & 0 \end{pmatrix}, \\
\lambda^7 =& \begin{pmatrix} 0 & 0 & 0 \\ 0 & 0 & -i \\ 0 & i & 0 \end{pmatrix} 
&\lambda^8 =& \frac{1}{\sqrt{3}}\begin{pmatrix} 1 & 0 & 0 \\ 0 & 1 & 0 \\ 0 & 0 & -2 \end{pmatrix}.
\end{align}

\section{Bose-Kondo Coupling \label{app:rkky}}
\begin{figure}[ht]
    \centering
    \includegraphics{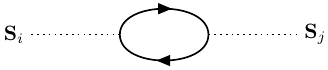}
    \caption{Effective Kondo Lattice RKKY Interaction; dotted lines refer to the pseudospin operators (not to be confused with dashed lines in other diagrams referring to pseudofermion propagators) and the solid lines are fermion propagators. The $\textbf{S}_i$ and $\textbf{S}_j$ are the spin operators on different sites $i$ and $j$ in the parent Kondo lattice.}
    \label{fig:rkky}
\end{figure}

In order to construct the coupling of the local moment to the bosonic bath while respecting the local symmetry, we construct the effective interaction between spins in the corresponding Kondo lattice. Starting with the Fermi-Kondo Hamiltonian in Eqs.~\eqref{eq:HQ1}-\eqref{eq:HO}, we can compute the effective interaction between two spins by computing the diagram in Fig.~\ref{fig:rkky}. We then replace one of the spin operators in this resulting RKKY interaction with the bosonic field and thereby find the symmetry-allowed coupling of the local moment to the bosonic bath. We emphasize that this is not an actual RKKY interaction between local moments on different sites, and should be conceptually likened to a Weiss mean field coupled to the impurity.

\section{Details of the renormalization group method \label{app:rg}}
\begin{figure}[H]
    \centering
    \includegraphics[scale = 0.8]{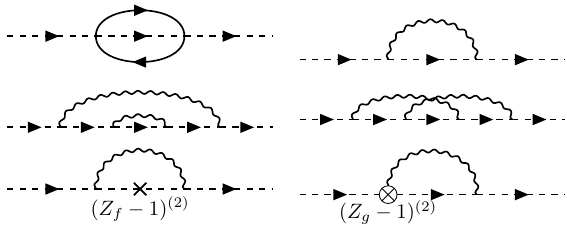}
    \caption{Pseudofermion self-energy, both direct and counterterm contribution.}
    \label{fig:pf_self_energy}
\end{figure}

\begin{figure}[H]
    \centering
    \includegraphics[scale = 0.8]{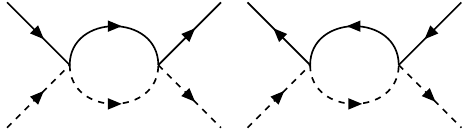}
    \caption{Order $K$ corrections to the Fermi-Kondo vertex. There are only direct contributions at this order.}
    \label{fig:J_1st_corr}
\end{figure}

\begin{figure}[H]
    \centering
    \includegraphics[scale = 0.8]{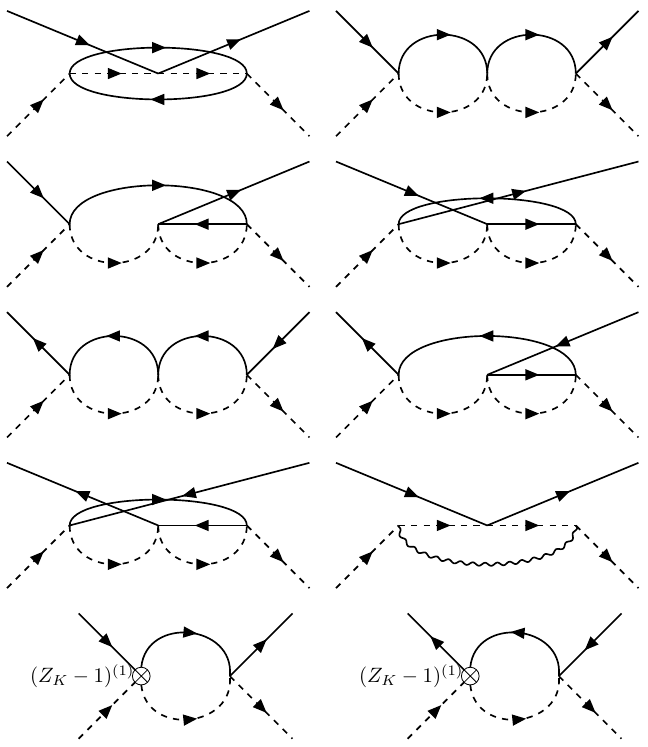}
    \caption{Order $K^2$ and $g^2$ corrections to the Fermi-Kondo vertex.}
    \label{fig:J_2nd_corr}
\end{figure}

\begin{figure}[H]
    \centering
    \includegraphics[scale = 0.8]{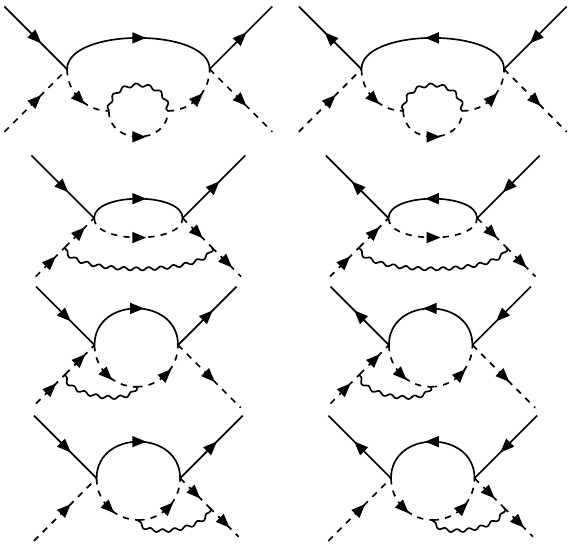}
    \caption{Order $Kg^2$ direct correction to the Fermi-Kondo vertex.}
    \label{fig:J_3rd_corr_direct}
\end{figure}

\begin{figure}[H]
    \centering
    \includegraphics[scale = 0.8]{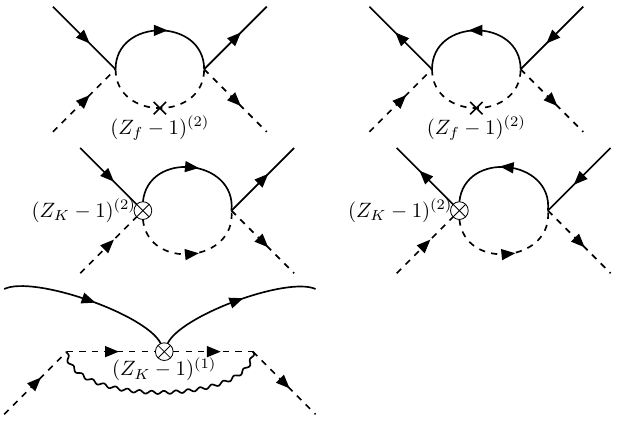}
    \caption{Order $Kg^2$ counterterm corrections to the Fermi-Kondo vertex.}
    \label{fig:J_3rd_corr_counter}
\end{figure}

\begin{figure}[H]
    \centering
    \includegraphics[scale = 0.8]{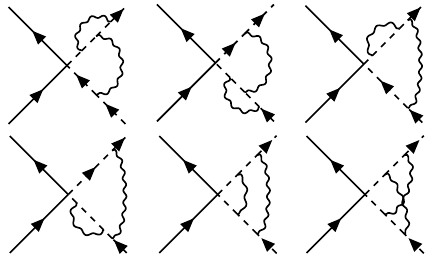}
    \caption{Order $g^4$ direct corrections to the Fermi-Kondo vertex.}
    \label{fig:J_4th_corr_direct}
\end{figure}

\begin{figure}[H]
    \centering
    \includegraphics[scale = 0.8]{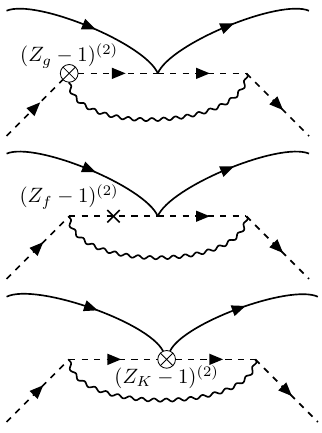}
    \caption{Order $g^4$ counterterm corrections to the Fermi-Kondo vertex.}
    \label{fig:J_4th_corr_counter}
\end{figure}

\begin{figure}[H]
    \centering
    \includegraphics[scale=0.8]{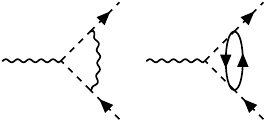}
    \caption{Order $g^2$ and $K^2$ corrections to the Bose-Kondo vertex.}
    \label{fig:g_2nd_corr}
\end{figure}

\begin{figure}[H]
    \centering
    \includegraphics[scale = 0.8]{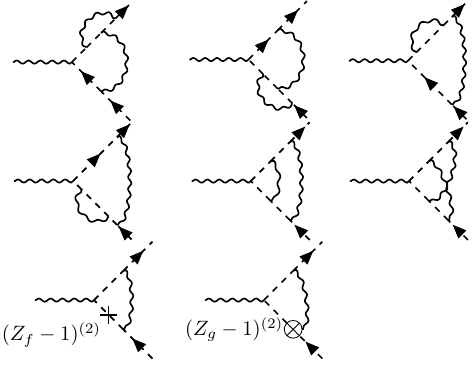}
    \caption{Order $g^4$ direct and counterterm corrections to the Bose-Kondo vertex.}
    \label{fig:g_4th_corr}
\end{figure}

From the bare Hamiltonian presented in the main text, we can introduce counterterms in order to remove divergences in the loop integrals. When calculating the Fermi-Kondo and Bose-Kondo vertex functions, as well as the pseudofermion self-energy, we can solve for these counterterms order by order and use them to compute the renormalization factors. The corresponding diagrams for the pseudofermion self-energy are given in Fig. \ref{fig:pf_self_energy}, the diagrams for the Fermi-Kondo vertex corrections are given in Figs. \ref{fig:J_1st_corr}-\ref{fig:J_4th_corr_counter}, and the diagrams for the Bose-Kondo vertex corrections are given in Figs. \ref{fig:g_2nd_corr}-\ref{fig:g_4th_corr}. Details for how to extract renormalization constants from the vertex corrections and self-energy are presented in an excellent reference \cite{Zhu2002}. In the Feynman diagrams of Figs.~\ref{fig:pf_self_energy}-\ref{fig:g_4th_corr}, solid lines refer to conduction electron propagators (Eq. \eqref{eq:fermion_gf}, dashed lines corresponds to pseudofermion propagators (Eq. \eqref{eq:pseudofermion_gf}), and the squiggly lines refer to bosonic bath propagators (Eq. \eqref{eq:boson_gf}):
\begin{align}
\mathcal{G}^c_0(i\omega,\mathbf{k}) ={}& \frac{1}{i\omega-E_{\mathbf{k}}} \label{eq:fermion_gf} \\
\mathcal{G}^f_0(i\omega,\mathbf{k}) ={}& \frac{1}{i\omega-\lambda} \label{eq:pseudofermion_gf} \\
\mathcal{G}^\phi_0(i\omega,\mathbf{k}) ={}& \frac{2\Omega_{\mathbf{k}}}{(i\omega-\Omega_{\mathbf{k}})(i\omega+\Omega_{\mathbf{k}})} \label{eq:boson_gf}
\end{align}

The renormalization constants and wavefunction renormalization up to third order in $K_{i}$ and fifth order in $g_{i}$, are given in Eqs.~\eqref{eq:ZKQ1}-\eqref{eq:Zf}. 
\begin{widetext}
\begin{align}
Z_{K_{Q1}}={}&
1
-\frac{1}{\epsilon_{Q}}\frac{g_{Q}^{2}(g_{Q}^{2}-g_{O}^{2})}{8}
+\frac{1}{\epsilon_{O}}\left[\frac{g_{O}^{2}}{4}+\frac{g_{O}^{2}g_{Q}^{2}}{8} \right]
+\frac{1}{\epsilon_{O}^{2}}\frac{g_{O}^{4}}{32}
+\frac{1}{\epsilon_{Q}\epsilon_{O}}\frac{3g_{Q}^{2}g_{O}^{2}}{8}
+\frac{1}{\epsilon_{O}\epsilon'}\frac{3K_{Q2}K_{O}g_{O}^{2}}{2K_{Q1}}\notag\\
&+\frac{1}{\epsilon'}\left[ \frac{3K_{O}^{2}}{2}+\frac{6K_{Q2}K_{O}}{K_{Q1}} \right]
+\frac{1}{\epsilon'^{2}}\left[
12K_{Q2}^{2}+3K_{O}^{2}-\frac{3\sqrt{3}K_{Q2}}{K_{Q1}}(2K_{Q2}^{2}+K_{O}^{2})
+\frac{9K_{Q2}K_{O}}{K_{Q1}}(4K_{Q2}^{2}+K_{O}^{2})
\right]\notag\\
&
-\frac{1}{\epsilon_{Q}(\epsilon_{Q}+\epsilon_{O})}\frac{g_{Q}^{2}g_{O}^{2}}{4}
+\frac{1}{\epsilon'(\epsilon_{Q}+\epsilon')}\frac{6K_{Q2}K_{O}g_{Q}^{2}}{K_{Q1}}, \label{eq:ZKQ1} \\
Z_{K_{Q2}}={}&
1
-\frac{1}{\epsilon_{Q}}\frac{g_{Q}^{2}(g_{Q}^{2}-g_{O}^{2})}{8}
+\frac{1}{\epsilon_{O}}\left[\frac{g_{O}^{2}}{4}+\frac{g_{O}^{2}g_{Q}^{2}}{8} \right]
+\frac{1}{\epsilon_{O}^{2}}\frac{g_{O}^{4}}{32}
+\frac{1}{\epsilon_{Q}\epsilon_{O}}\frac{3g_{Q}^{2}g_{O}^{2}}{8}
+\frac{1}{\epsilon_{O}\epsilon'}\left[ \frac{K_{Q1}K_{O}g_{O}^{2}}{4K_{Q2}} -\frac{\sqrt{3}K_{O}g_{O}^{2}}{4} \right]\notag\\
&+\frac{1}{\epsilon'}\left[ \frac{K_{Q1}K_{O}}{K_{Q2}}-\sqrt{3}K_{O}+\frac{3K_{O}^{2}}{2} \right]\notag\\
&+\frac{1}{\epsilon'^{2}}\left[
2K_{Q1}^{2}+3K_{Q2}^{2}+\frac{9K_{O}^{2}}{2}-3\sqrt{3}K_{Q1}K_{Q2}-\frac{\sqrt{3}K_{Q1}K_{O}^{2}}{K_{Q2}}
+K_{O}\left(-\sqrt{3}K_{Q1}^{2}+6K_{Q1}K_{Q2}-6\sqrt{3}K_{Q2}^{2}-\frac{3\sqrt{3}K_{O}^{2}}{2} \right)\right.\notag\\
&\quad\quad\quad\left.+\frac{K_{Q1}K_{O}}{2K_{Q2}}(2K_{Q1}^{2}+3K_{O}^{2})
\right]
-\frac{1}{\epsilon_{Q}(\epsilon_{Q}+\epsilon_{O})}\frac{g_{Q}^{2}g_{O}^{2}}{4}
+\frac{1}{\epsilon'(\epsilon_{Q}+\epsilon')}\left[ \frac{K_{Q1}K_{O}g_{Q}^{2}}{K_{Q2}}-\sqrt{3}K_{O}g_{Q}^{2}\right], \label{eq:ZKQ2} \\
Z_{K_{O}}={}&1
+\frac{1}{\epsilon_{Q}}\left[ \frac{g_{Q}^{2}}{2}+\frac{g_{Q}^{4}}{8}+\frac{g_{Q}^{2}g_{O}^{2}}{4}\right]
+\frac{1}{\epsilon_{Q}^{2}}\frac{3g_{Q}^{4}}{8}
-\frac{1}{\epsilon_{O}}\frac{g_{O}^{2}+g_{Q}^{2}g_{O}^{2}}{4}
+\frac{1}{\epsilon_{O}^{2}}\frac{g_{O}^{4}}{32}\notag\\
&-\frac{1}{\epsilon_{Q}\epsilon_{O}}\frac{3g_{Q}^{2}g_{O}^{2}}{8}
+\frac{1}{\epsilon_{O}\epsilon'}\left[ -\frac{K_{Q1}K_{Q2}g_{O}^{2}}{K_{O}}+\frac{\sqrt{3}K_{Q2}^{2}g_{O}^{2}}{K_{O}} \right]
+\frac{1}{\epsilon_{Q}\epsilon'}\left[ \frac{2K_{Q1}K_{Q2}g_{Q}^{2}}{K_{O}}-\frac{\sqrt{3}K_{Q2}^{2}g_{Q}^{2}}{K_{O}} \right]\notag\\
&+\frac{1}{\epsilon'}\left[ K_{Q1}^{2}+6K_{Q2}^{2}-\frac{3K_{O}^{2}}{2}+\frac{2K_{Q2}}{K_{O}}(2K_{Q2}-\sqrt{3}K_{Q2}) \right]\notag\\
&+\frac{1}{\epsilon'^{2}}\left[
2K_{Q1}^{2}-4\sqrt{3}K_{Q1}K_{Q2}+18K_{Q2}^{2}
+3K_{Q2}K_{O}(2K_{Q1}-\sqrt{2}K_{Q2})
+\frac{2K_{Q2}}{K_{O}}(K_{Q1}^{2}+6K_{Q2}^{2})(2K_{Q1}-\sqrt{3}K_{Q2})
\right]\notag\\
&-\frac{1}{\epsilon_{Q}(\epsilon_{Q}+\epsilon_{O})}\frac{g_{Q}^{2}g_{O}^{2}}{4}
+\frac{1}{\epsilon'(\epsilon_{Q}+\epsilon')}\left[ \frac{K_{Q1}K_{O}g_{Q}^{2}}{K_{Q2}}-\sqrt{3}K_{O}g_{Q}^{2} \right],\label{eq:ZKO} \\
Z_{g_{Q}}={}&1
-\frac{1}{\epsilon_{Q}}\frac{g_{Q}^{2}(g_{Q}^{2}+g_{O}^{2})}{8}
+\frac{1}{\epsilon_{O}}\left[ \frac{g_{O}^{2}}{4}+\frac{g_{Q}^{2}g_{O}^{2}}{8}\right]
+\frac{1}{\epsilon_{O}^{2}}\frac{g_{O}^{4}}{32}
+\frac{1}{\epsilon'}\frac{3K_{O}^{2}}{2}
+\frac{1}{\epsilon_{Q}\epsilon_{O}}\frac{3g_{Q}^{2}g_{O}^{2}}{8}
-\frac{1}{\epsilon_{Q}(\epsilon_{Q}+\epsilon_{O})}\frac{g_{Q}^{2}g_{O}^{2}}{4}, \label{eq:ZgQ} \\
Z_{g_{O}}={}&1
+\frac{1}{\epsilon_{Q}}\left[ \frac{g_{Q}^{2}}{2}+\frac{g_{Q}^{4}}{8}+\frac{g_{Q}^{2}g_{O}^{2}}{4} \right]
+\frac{1}{\epsilon_{Q}^{2}}\frac{3g_{Q}^{4}}{8}
-\frac{1}{\epsilon_{O}}\frac{(g_{O}^{2}+g_{Q}^{2}g_{O}^{2})}{4}
+\frac{1}{\epsilon_{O}^{2}}\frac{g_{O}^{4}}{32}
+\frac{1}{\epsilon'}\left[K_{Q1}^{2}+6K_{Q2}^{2}-\frac{3K_{O}^{2}}{2}\right]\notag\\
&-\frac{1}{\epsilon_{Q}\epsilon_{O}}\frac{3g_{Q}^{2}g_{O}^{2}}{8}
-\frac{1}{(\epsilon_{Q}+\epsilon_{O})}\frac{g_{Q}^{2}g_{O}^{2}}{2}
+\frac{1}{(\epsilon_{Q}+\epsilon_{O})}\frac{g_{Q}^{2}g_{O}^{2}}{2}, \label{eq:ZgO} \\
Z_{f}={}&
1
+\frac{1}{\epsilon_{Q}}\left[ -\frac{g_{Q}^{2}}{2}+\frac{g_{Q}^{4}}{8}\right]
-\frac{1}{\epsilon_{Q}^{2}}\frac{g_{Q}^{4}}{8}
-\frac{1}{\epsilon_{O}}\frac{g_{O}^{2}}{4}
+\frac{1}{\epsilon_{O}^{2}}\frac{g_{O}^{4}}{32}
-\frac{1}{\epsilon'}\left[ K_{Q1}^{2}+6K_{Q2}^{2}+\frac{3K_{O}^{2}}{2} \right]+\frac{1}{(\epsilon_{Q}+\epsilon_{O})}\frac{g_{Q}^{2}g_{O}^{2}}{2}-\frac{1}{\epsilon_{Q}\epsilon_{O}}\frac{3g_{Q}^{2}g_{O}^{2}}{8}.\label{eq:Zf}
\end{align}
From the renormalization constants, we can compute the beta functions,
\begin{align}
\frac{d K_{i}}{d\ln\mu}={}&K_{i}\left[\sum_{k=Q_{1},Q_{2},O}K_{k}\partial_{K_{k}}G_{K_{i}}^{(0,0,1)}+\frac{g_{Q}}{2}\partial_{g_{Q}}G_{K_{i}}^{(1,0,0)}+\frac{g_{O}}{2}\partial_{g_{O}}G_{K_{i}}^{(0,1,0)}\right], \label{eq:K_beta}\\
\frac{d g_{j}}{d\ln\mu}={}&g_{j}\left[-\frac{\epsilon_{j}}{2}+\sum_{k=Q_{1},Q_{2},O}K_{k}\partial_{K_{k}}G_{g_{j}}^{(0,0,1)}+\frac{g_{Q}}{2}\partial_{g_{Q}}G_{g_{j}}^{(1,0,0)}+\frac{g_{O}}{2}\partial_{g_{O}}G_{g_{j}}^{(0,1,0)}\right], \label{eq:g_beta} 
\end{align}
\end{widetext}
\noindent where we Taylor expand the products $Z^{-1}_{f}Z_{K_i}$ and $Z^{-1}_{f}Z_{g_j}$ as follows in order to obtain the $G^{(m,n,\ell)}$ factors which appear in Eqs.~\eqref{eq:K_beta},\eqref{eq:g_beta}:
\begin{align}
G_{K_{i}}\equiv Z_{f}^{-1}Z_{K_{i}}={}&\sum_{m,n,\ell=0}^{\infty}\frac{G_{K_{i}}^{(m,n,\ell)}(\{K,g\})}{\epsilon_{Q}^{m}\epsilon_{O}^{n}\epsilon'^{\ell}}, \\
G_{g_{j}}\equiv Z_{f}^{-1}Z_{g_{j}}={}&\sum_{m,n,\ell=0}^{\infty}\frac{G_{g_{j}}^{(m,n,\ell)}(\{K,g\})}{\epsilon_{Q}^{m}\epsilon_{O}^{n}\epsilon'^{\ell}}.
\end{align}
The first terms of the series are $G_{K_{i}}^{(0,0,0)}=G_{g_{j}}^{(0,0,0)}=1$, and the indices $i=Q_{1},Q_{2},O$, $j=Q,O$.

\section{Exponents for multipolar susceptibility \label{app:multipolar_susc}}

From the Bose-Kondo beta functions Eq.~\eqref{eq:beta_lambdaQ}-\eqref{eq:beta_lambdaO} in Sec.~\ref{sec:bfkondo}, the local multipolar moment susceptibilities for the case of zero fixed point values in our model are given by Eq.~\eqref{eq:suscep_zero}, and turn out to be
\begin{align}
\gamma_{Q}={}&\lambda_{O}^{*}+4((K_{Q1}^{*})^{2}+6(K_{Q2}^{*})^{2}+3(K_{O}^{*})^{2}),
\end{align}
if $\lambda_{Q}^{*}=0$, and
\begin{align}
\gamma_{O}={}&2\lambda_{Q}^{*}+8((K_{Q1}^{*})^{2}+6(K_{Q2}^{*})^{2}),
\end{align}
if $\lambda_{O}^{*}=0$. The latter of these two is used to compute the octupolar susceptibility exponent at the critical points $C^{Q}_{1,2\pm}$.

\section{Scaling behaviors of multipolar susceptibility at finite temperature}\label{app:temp_suscep}
The scaling behavior of the multipolar susceptibility as a function of imaginary time in the previous section is for zero temperature. Here, we will discuss how to obtain the scaling behavior for finite temperature. 
Let us assume that we have conformal invariance at the critical point. Assuming that the multipolar moments are primary operators with conformal dimension $\gamma_{i}/2$, the correlation function (susceptibility) of the multipolar moment is \cite{Aronson1997,Zarand2002}
\begin{align}
\braket{T_{\tau}S^{i}(\tau_{1})S^{i}(\tau_{2})}\propto{}&\frac{1}{|\tau_{1}-\tau_{2}|^{\gamma_{i}}}.
\end{align}
Performing a conformal mapping, $\tau\rightarrow f(\tau)=\frac{\pi}{\beta}\tan\left(\frac{\pi \tau}{\beta}\right)$,
\begin{align}
&\braket{T_{\tau}S^{i}(\tau_{1})S^{i}(\tau_{2})}\rightarrow\notag\\ &\left(\frac{\partial f(\tau_{1})}{\partial\tau_{1}}\right)^{\gamma_{i}/2} \left(\frac{\partial f(\tau_{2})}{\partial\tau_{2}}\right)^{\gamma_{i}/2} \braket{T_{\tau}S^{i}(f(\tau_{1}))S^{i}(f(\tau_{2}))}.
\end{align}
Letting $\tau_{1}=\tau$ and $\tau_{2}=0$, then
\begin{align}
\chi_{i}(\tau,T)\propto \left(\frac{\pi/\beta}{\sin(\pi\tau/\beta)}\right)^{\gamma_{i}}.
\end{align}
After Fourier transforming and analytic continuation, we can get the multipolar susceptibility in terms of the temperature $T$ and the energy scaling $\omega$ \cite{Parcollet1998,Parcollet1999,Aronson1997},
\begin{align}
\chi_{i}(\omega,T)\propto{}&T^{\gamma_{i}-1}\frac{\Gamma(\frac{\gamma_{i}}{2}-\frac{i\omega}{2\pi T})\Gamma(1-\gamma_{i})}{\Gamma(1-\frac{\gamma_{i}}{2}-\frac{i\omega}{2\pi T})}.
\end{align}
The scaling behavior of the real and imaginary parts of  $F(\frac{\omega}{T})\equiv\Gamma(\frac{\gamma_{i}}{2}-\frac{i\omega}{2\pi T})/\Gamma(1-\frac{\gamma_{i}}{2}-\frac{i\omega}{2\pi T})$ is
\begin{align}
        \text{Re}[F(x)]={}&\begin{cases}
    F(0)+C_{\text{Re},<}|x|^{2},&|x|\ll1,\\
    C_{\text{Re},>}|x|^{\gamma_{i}-1},&|x|\gg1,\\
    \end{cases}\\
    \text{Im}[F(x)]={}&\begin{cases}
    C_{\text{Im},<}|x|,&|x|\ll1,\\
    C_{\text{Im},>}|x|^{\gamma_{i}-1},&|x|\gg1,\\
    \end{cases},
\end{align}
where $F(0)=\frac{\Gamma(\frac{\gamma_{i}}{2})\Gamma(1-\gamma_{i})}{\Gamma(1-\frac{\gamma_{i}}{2})}$ and $C_{\text{Re},<}$ is a real constant.
Then, the temperature dependencies of the real and imaginary parts of the multipolar susceptibility are given in Eqs.~\eqref{eq:real_susc}, \eqref{eq:imag_susc}, where the constant $C_{\text{Re}1}=C_{\text{Re},<}/F(0)$ in the main text.
Note that the real part of the susceptibility exponent at the dc limit ($\omega=0$) for the Kondo fixed point is consistent with the CFT result \cite{Affleck1993b}, $\chi'\sim T^{2\Delta-1}$, where $\gamma_{i}=2\Delta$.


\begin{thebibliography}{92}%
\makeatletter
\providecommand \@ifxundefined [1]{%
 \@ifx{#1\undefined}
}%
\providecommand \@ifnum [1]{%
 \ifnum #1\expandafter \@firstoftwo
 \else \expandafter \@secondoftwo
 \fi
}%
\providecommand \@ifx [1]{%
 \ifx #1\expandafter \@firstoftwo
 \else \expandafter \@secondoftwo
 \fi
}%
\providecommand \natexlab [1]{#1}%
\providecommand \enquote  [1]{``#1''}%
\providecommand \bibnamefont  [1]{#1}%
\providecommand \bibfnamefont [1]{#1}%
\providecommand \citenamefont [1]{#1}%
\providecommand \href@noop [0]{\@secondoftwo}%
\providecommand \href [0]{\begingroup \@sanitize@url \@href}%
\providecommand \@href[1]{\@@startlink{#1}\@@href}%
\providecommand \@@href[1]{\endgroup#1\@@endlink}%
\providecommand \@sanitize@url [0]{\catcode `\\12\catcode `\$12\catcode
  `\&12\catcode `\#12\catcode `\^12\catcode `\_12\catcode `\%12\relax}%
\providecommand \@@startlink[1]{}%
\providecommand \@@endlink[0]{}%
\providecommand \url  [0]{\begingroup\@sanitize@url \@url }%
\providecommand \@url [1]{\endgroup\@href {#1}{\urlprefix }}%
\providecommand \urlprefix  [0]{URL }%
\providecommand \Eprint [0]{\href }%
\providecommand \doibase [0]{http://dx.doi.org/}%
\providecommand \selectlanguage [0]{\@gobble}%
\providecommand \bibinfo  [0]{\@secondoftwo}%
\providecommand \bibfield  [0]{\@secondoftwo}%
\providecommand \translation [1]{[#1]}%
\providecommand \BibitemOpen [0]{}%
\providecommand \bibitemStop [0]{}%
\providecommand \bibitemNoStop [0]{.\EOS\space}%
\providecommand \EOS [0]{\spacefactor3000\relax}%
\providecommand \BibitemShut  [1]{\csname bibitem#1\endcsname}%
\let\auto@bib@innerbib\@empty
\bibitem [{\citenamefont {Kusunose}(2008)}]{Kusunose2008b}%
  \BibitemOpen
  \bibfield  {author} {\bibinfo {author} {\bibfnamefont {H.}~\bibnamefont
  {Kusunose}},\ }\href {\doibase 10.1143/JPSJ.77.064710} {\bibfield  {journal}
  {\bibinfo  {journal} {Journal of the Physical Society of Japan}\ }\textbf
  {\bibinfo {volume} {77}},\ \bibinfo {pages} {064710} (\bibinfo {year}
  {2008})}\BibitemShut {NoStop}%
\bibitem [{\citenamefont {Kuramoto}\ \emph {et~al.}(2009)\citenamefont
  {Kuramoto}, \citenamefont {Kusunose},\ and\ \citenamefont
  {Kiss}}]{Kuramoto2009a}%
  \BibitemOpen
  \bibfield  {author} {\bibinfo {author} {\bibfnamefont {Y.}~\bibnamefont
  {Kuramoto}}, \bibinfo {author} {\bibfnamefont {H.}~\bibnamefont {Kusunose}},
  \ and\ \bibinfo {author} {\bibfnamefont {A.}~\bibnamefont {Kiss}},\ }\href
  {\doibase 10.1143/JPSJ.78.072001} {\bibfield  {journal} {\bibinfo  {journal}
  {Journal of the Physical Society of Japan}\ }\textbf {\bibinfo {volume}
  {78}},\ \bibinfo {pages} {072001} (\bibinfo {year} {2009})}\BibitemShut
  {NoStop}%
\bibitem [{\citenamefont {Martelli}\ \emph {et~al.}(2019)\citenamefont
  {Martelli}, \citenamefont {Cai}, \citenamefont {Nica}, \citenamefont
  {Taupin}, \citenamefont {Prokofiev}, \citenamefont {Liu}, \citenamefont
  {Lai}, \citenamefont {Yu}, \citenamefont {Ingersent}, \citenamefont
  {K{\"{u}}chler}, \citenamefont {Strydom}, \citenamefont {Geiger},
  \citenamefont {Haenel}, \citenamefont {Larrea}, \citenamefont {Si},\ and\
  \citenamefont {Paschen}}]{Martelli2019}%
  \BibitemOpen
  \bibfield  {author} {\bibinfo {author} {\bibfnamefont {V.}~\bibnamefont
  {Martelli}}, \bibinfo {author} {\bibfnamefont {A.}~\bibnamefont {Cai}},
  \bibinfo {author} {\bibfnamefont {E.~M.}\ \bibnamefont {Nica}}, \bibinfo
  {author} {\bibfnamefont {M.}~\bibnamefont {Taupin}}, \bibinfo {author}
  {\bibfnamefont {A.}~\bibnamefont {Prokofiev}}, \bibinfo {author}
  {\bibfnamefont {C.~C.}\ \bibnamefont {Liu}}, \bibinfo {author} {\bibfnamefont
  {H.~H.}\ \bibnamefont {Lai}}, \bibinfo {author} {\bibfnamefont
  {R.}~\bibnamefont {Yu}}, \bibinfo {author} {\bibfnamefont {K.}~\bibnamefont
  {Ingersent}}, \bibinfo {author} {\bibfnamefont {R.}~\bibnamefont
  {K{\"{u}}chler}}, \bibinfo {author} {\bibfnamefont {A.~M.}\ \bibnamefont
  {Strydom}}, \bibinfo {author} {\bibfnamefont {D.}~\bibnamefont {Geiger}},
  \bibinfo {author} {\bibfnamefont {J.}~\bibnamefont {Haenel}}, \bibinfo
  {author} {\bibfnamefont {J.}~\bibnamefont {Larrea}}, \bibinfo {author}
  {\bibfnamefont {Q.}~\bibnamefont {Si}}, \ and\ \bibinfo {author}
  {\bibfnamefont {S.}~\bibnamefont {Paschen}},\ }\href {\doibase 10.1073/pnas.1908101116} {\bibfield  {journal} {\bibinfo  {journal}
  {Proceedings of the National Academy of Sciences of the United States of
  America}\ }\textbf {\bibinfo {volume} {116}},\ \bibinfo {pages} {17701}
  (\bibinfo {year} {2019})}\BibitemShut {NoStop}%
\bibitem [{\citenamefont {Rosenberg}\ \emph {et~al.}(2019)\citenamefont
  {Rosenberg}, \citenamefont {Chu}, \citenamefont {Ruff}, \citenamefont
  {Hristov},\ and\ \citenamefont {Fisher}}]{Rosenberg2019c}%
  \BibitemOpen
  \bibfield  {author} {\bibinfo {author} {\bibfnamefont {E.~W.}\ \bibnamefont
  {Rosenberg}}, \bibinfo {author} {\bibfnamefont {J.~H.}\ \bibnamefont {Chu}},
  \bibinfo {author} {\bibfnamefont {J.~P.}\ \bibnamefont {Ruff}}, \bibinfo
  {author} {\bibfnamefont {A.~T.}\ \bibnamefont {Hristov}}, \ and\ \bibinfo
  {author} {\bibfnamefont {I.~R.}\ \bibnamefont {Fisher}},\ }\href {\doibase 10.1073/pnas.1818910116} {\bibfield  {journal} {\bibinfo  {journal}
  {Proceedings of the National Academy of Sciences of the United States of
  America}\ }\textbf {\bibinfo {volume} {116}},\ \bibinfo {pages} {7232}
  (\bibinfo {year} {2019})}\BibitemShut {NoStop}%
\bibitem [{\citenamefont {White}\ \emph {et~al.}(2015)\citenamefont {White},
  \citenamefont {Thompson},\ and\ \citenamefont {Maple}}]{White2015}%
  \BibitemOpen
  \bibfield  {author} {\bibinfo {author} {\bibfnamefont {B.~D.}\ \bibnamefont
  {White}}, \bibinfo {author} {\bibfnamefont {J.~D.}\ \bibnamefont {Thompson}},
  \ and\ \bibinfo {author} {\bibfnamefont {M.~B.}\ \bibnamefont {Maple}},\
  }\href {\doibase 10.1016/j.physc.2015.02.044} {\bibfield  {journal} {\bibinfo
   {journal} {Physica C: Superconductivity and its Applications}\ }\textbf
  {\bibinfo {volume} {514}},\ \bibinfo {pages} {246} (\bibinfo {year}
  {2015})}\BibitemShut {NoStop}%
\bibitem [{\citenamefont {{Saxena, S S. Agarwal, P. Ahilan, K. Grosche, F M. W
  Haselwimmer, R K. Steiner, M J. Pugh, E. Walker, I R. Julian, S R. Monthoux,
  P. Lonzarich, G G. Huxley, A. Sheikin, I. Braithwaite, D.
  Flouquet}}(2000)}]{Saxena2000}%
  \BibitemOpen
  \bibfield  {author} {\bibinfo {author} {\bibfnamefont {J.}~\bibnamefont
  {{Saxena, S S. Agarwal, P. Ahilan, K. Grosche, F M. W Haselwimmer, R K.
  Steiner, M J. Pugh, E. Walker, I R. Julian, S R. Monthoux, P. Lonzarich, G G.
  Huxley, A. Sheikin, I. Braithwaite, D. Flouquet}}},\ }\href {www.nature.com}
  {\bibfield  {journal} {\bibinfo  {journal} {Nature}\ }\textbf {\bibinfo
  {volume} {406}},\ \bibinfo {pages} {587} (\bibinfo {year}
  {2000})}\BibitemShut {NoStop}%
\bibitem [{\citenamefont {Kohori}\ \emph {et~al.}(2000)\citenamefont {Kohori},
  \citenamefont {Yamato}, \citenamefont {Iwamoto},\ and\ \citenamefont
  {Kohara}}]{Kohori2000}%
  \BibitemOpen
  \bibfield  {author} {\bibinfo {author} {\bibfnamefont {Y.}~\bibnamefont
  {Kohori}}, \bibinfo {author} {\bibfnamefont {Y.}~\bibnamefont {Yamato}},
  \bibinfo {author} {\bibfnamefont {Y.}~\bibnamefont {Iwamoto}}, \ and\
  \bibinfo {author} {\bibfnamefont {T.}~\bibnamefont {Kohara}},\ }\href
  {\doibase 10.1007/PL00011077} {\bibfield  {journal} {\bibinfo  {journal}
  {European Physical Journal B}\ }\textbf {\bibinfo {volume} {18}},\ \bibinfo
  {pages} {601} (\bibinfo {year} {2000})}\BibitemShut {NoStop}%
\bibitem [{\citenamefont {Izawa}\ \emph {et~al.}(2001)\citenamefont {Izawa},
  \citenamefont {Yamaguchi}, \citenamefont {Matsuda}, \citenamefont {Shishido},
  \citenamefont {Settai},\ and\ \citenamefont {Onuki}}]{Izawa2001}%
  \BibitemOpen
  \bibfield  {author} {\bibinfo {author} {\bibfnamefont {K.}~\bibnamefont
  {Izawa}}, \bibinfo {author} {\bibfnamefont {H.}~\bibnamefont {Yamaguchi}},
  \bibinfo {author} {\bibfnamefont {Y.}~\bibnamefont {Matsuda}}, \bibinfo
  {author} {\bibfnamefont {H.}~\bibnamefont {Shishido}}, \bibinfo {author}
  {\bibfnamefont {R.}~\bibnamefont {Settai}}, \ and\ \bibinfo {author}
  {\bibfnamefont {Y.}~\bibnamefont {Onuki}},\ }\href {\doibase 10.1103/PhysRevLett.87.057002} {\bibfield  {journal} {\bibinfo  {journal}
  {Physical Review Letters}\ }\textbf {\bibinfo {volume} {87}},\ \bibinfo
  {pages} {57002} (\bibinfo {year} {2001})}\BibitemShut {NoStop}%
\bibitem [{\citenamefont {Aoki}\ \emph {et~al.}(2003)\citenamefont {Aoki},
  \citenamefont {Tsuchiya}, \citenamefont {Kanayama}, \citenamefont {Saha},
  \citenamefont {Sugawara}, \citenamefont {Sato}, \citenamefont {Higemoto},
  \citenamefont {Koda}, \citenamefont {Ohishi}, \citenamefont {Nishiyama},\
  and\ \citenamefont {Kadono}}]{Aoki2003}%
  \BibitemOpen
  \bibfield  {author} {\bibinfo {author} {\bibfnamefont {Y.}~\bibnamefont
  {Aoki}}, \bibinfo {author} {\bibfnamefont {A.}~\bibnamefont {Tsuchiya}},
  \bibinfo {author} {\bibfnamefont {T.}~\bibnamefont {Kanayama}}, \bibinfo
  {author} {\bibfnamefont {S.~R.}\ \bibnamefont {Saha}}, \bibinfo {author}
  {\bibfnamefont {H.}~\bibnamefont {Sugawara}}, \bibinfo {author}
  {\bibfnamefont {H.}~\bibnamefont {Sato}}, \bibinfo {author} {\bibfnamefont
  {W.}~\bibnamefont {Higemoto}}, \bibinfo {author} {\bibfnamefont
  {A.}~\bibnamefont {Koda}}, \bibinfo {author} {\bibfnamefont {K.}~\bibnamefont
  {Ohishi}}, \bibinfo {author} {\bibfnamefont {K.}~\bibnamefont {Nishiyama}}, \
  and\ \bibinfo {author} {\bibfnamefont {R.}~\bibnamefont {Kadono}},\ }\href
  {\doibase 10.1103/PhysRevLett.91.067003} {\bibfield  {journal} {\bibinfo
  {journal} {Physical Review Letters}\ }\textbf {\bibinfo {volume} {91}},\
  \bibinfo {pages} {067003} (\bibinfo {year} {2003})}\BibitemShut {NoStop}%
\bibitem [{\citenamefont {Bauer}\ \emph {et~al.}(2004)\citenamefont {Bauer},
  \citenamefont {Hilscher}, \citenamefont {Michor}, \citenamefont {Paul},
  \citenamefont {Scheidt}, \citenamefont {Gribanov}, \citenamefont {Seropegin},
  \citenamefont {No{\"{e}}l}, \citenamefont {Sigrist},\ and\ \citenamefont
  {Rogl}}]{Bauer2004}%
  \BibitemOpen
  \bibfield  {author} {\bibinfo {author} {\bibfnamefont {E.}~\bibnamefont
  {Bauer}}, \bibinfo {author} {\bibfnamefont {G.}~\bibnamefont {Hilscher}},
  \bibinfo {author} {\bibfnamefont {H.}~\bibnamefont {Michor}}, \bibinfo
  {author} {\bibfnamefont {C.}~\bibnamefont {Paul}}, \bibinfo {author}
  {\bibfnamefont {E.~W.}\ \bibnamefont {Scheidt}}, \bibinfo {author}
  {\bibfnamefont {A.}~\bibnamefont {Gribanov}}, \bibinfo {author}
  {\bibfnamefont {Y.}~\bibnamefont {Seropegin}}, \bibinfo {author}
  {\bibfnamefont {H.}~\bibnamefont {No{\"{e}}l}}, \bibinfo {author}
  {\bibfnamefont {M.}~\bibnamefont {Sigrist}}, \ and\ \bibinfo {author}
  {\bibfnamefont {P.}~\bibnamefont {Rogl}},\ }\href {\doibase 10.1103/PhysRevLett.92.027003} {\bibfield  {journal} {\bibinfo  {journal}
  {Physical Review Letters}\ }\textbf {\bibinfo {volume} {92}},\ \bibinfo
  {pages} {027003} (\bibinfo {year} {2004})}\BibitemShut {NoStop}%
\bibitem [{\citenamefont {Cox}(1987)}]{Cox1987a}%
  \BibitemOpen
  \bibfield  {author} {\bibinfo {author} {\bibfnamefont {D.~L.}\ \bibnamefont
  {Cox}},\ }\href {\doibase 10.1103/PhysRevLett.59.1240} {\bibfield  {journal}
  {\bibinfo  {journal} {Physical Review Letters}\ }\textbf {\bibinfo {volume}
  {59}},\ \bibinfo {pages} {1240} (\bibinfo {year} {1987})}\BibitemShut
  {NoStop}%
\bibitem [{\citenamefont {Jiao}\ \emph {et~al.}(2015)\citenamefont {Jiao},
  \citenamefont {Chen}, \citenamefont {Kohama}, \citenamefont {Graf},
  \citenamefont {Bauer}, \citenamefont {Singleton}, \citenamefont {Zhu},
  \citenamefont {Weng}, \citenamefont {Pang}, \citenamefont {Shang},
  \citenamefont {Zhang}, \citenamefont {Lee}, \citenamefont {Park},
  \citenamefont {Jaime}, \citenamefont {Thompson}, \citenamefont {Steglich},
  \citenamefont {Si},\ and\ \citenamefont {Yuan}}]{Jiao2015}%
  \BibitemOpen
  \bibfield  {author} {\bibinfo {author} {\bibfnamefont {L.}~\bibnamefont
  {Jiao}}, \bibinfo {author} {\bibfnamefont {Y.}~\bibnamefont {Chen}}, \bibinfo
  {author} {\bibfnamefont {Y.}~\bibnamefont {Kohama}}, \bibinfo {author}
  {\bibfnamefont {D.}~\bibnamefont {Graf}}, \bibinfo {author} {\bibfnamefont
  {E.~D.}\ \bibnamefont {Bauer}}, \bibinfo {author} {\bibfnamefont
  {J.}~\bibnamefont {Singleton}}, \bibinfo {author} {\bibfnamefont {J.~X.}\
  \bibnamefont {Zhu}}, \bibinfo {author} {\bibfnamefont {Z.}~\bibnamefont
  {Weng}}, \bibinfo {author} {\bibfnamefont {G.}~\bibnamefont {Pang}}, \bibinfo
  {author} {\bibfnamefont {T.}~\bibnamefont {Shang}}, \bibinfo {author}
  {\bibfnamefont {J.}~\bibnamefont {Zhang}}, \bibinfo {author} {\bibfnamefont
  {H.~O.}\ \bibnamefont {Lee}}, \bibinfo {author} {\bibfnamefont
  {T.}~\bibnamefont {Park}}, \bibinfo {author} {\bibfnamefont {M.}~\bibnamefont
  {Jaime}}, \bibinfo {author} {\bibfnamefont {J.~D.}\ \bibnamefont {Thompson}},
  \bibinfo {author} {\bibfnamefont {F.}~\bibnamefont {Steglich}}, \bibinfo
  {author} {\bibfnamefont {Q.}~\bibnamefont {Si}}, \ and\ \bibinfo {author}
  {\bibfnamefont {H.~Q.}\ \bibnamefont {Yuan}},\ }\href {\doibase 10.1073/pnas.1413932112} {\bibfield  {journal} {\bibinfo  {journal}
  {Proceedings of the National Academy of Sciences of the United States of
  America}\ }\textbf {\bibinfo {volume} {112}},\ \bibinfo {pages} {673}
  (\bibinfo {year} {2015})}\BibitemShut {NoStop}%
\bibitem [{\citenamefont {Kratochv{\'i}lov{\'a}}\ \emph
  {et~al.}(2015)\citenamefont {Kratochv{\'i}lov{\'a}}, \citenamefont
  {Prokle{\v{s}}ka}, \citenamefont {Uhl{\'i}{\v{r}}ov{\'a}}, \citenamefont
  {Tk{\'a}{\v{c}}}, \citenamefont {Du{\v{s}}ek}, \citenamefont
  {Sechovsk{\'y}},\ and\ \citenamefont {Custers}}]{Kratochvilova2015}%
  \BibitemOpen
  \bibfield  {author} {\bibinfo {author} {\bibfnamefont {M.}~\bibnamefont
  {Kratochv{\'i}lov{\'a}}}, \bibinfo {author} {\bibfnamefont {J.}~\bibnamefont
  {Prokle{\v{s}}ka}}, \bibinfo {author} {\bibfnamefont {K.}~\bibnamefont
  {Uhl{\'i}{\v{r}}ov{\'a}}}, \bibinfo {author} {\bibfnamefont {V.}~\bibnamefont
  {Tk{\'a}{\v{c}}}}, \bibinfo {author} {\bibfnamefont {M.}~\bibnamefont
  {Du{\v{s}}ek}}, \bibinfo {author} {\bibfnamefont {V.}~\bibnamefont
  {Sechovsk{\'y}}}, \ and\ \bibinfo {author} {\bibfnamefont {J.}~\bibnamefont
  {Custers}},\ }\href {\doibase 10.1038/srep15904} {\bibfield  {journal}
  {\bibinfo  {journal} {Scientific Reports}\ }\textbf {\bibinfo {volume} {5}},\
  \bibinfo {pages} {15904} (\bibinfo {year} {2015})}\BibitemShut {NoStop}%
\bibitem [{\citenamefont {Custers}\ \emph {et~al.}(2010)\citenamefont
  {Custers}, \citenamefont {Gegenwart}, \citenamefont {Geibel}, \citenamefont
  {Steglich}, \citenamefont {Coleman},\ and\ \citenamefont
  {Paschen}}]{Custers2010a}%
  \BibitemOpen
  \bibfield  {author} {\bibinfo {author} {\bibfnamefont {J.}~\bibnamefont
  {Custers}}, \bibinfo {author} {\bibfnamefont {P.}~\bibnamefont {Gegenwart}},
  \bibinfo {author} {\bibfnamefont {C.}~\bibnamefont {Geibel}}, \bibinfo
  {author} {\bibfnamefont {F.}~\bibnamefont {Steglich}}, \bibinfo {author}
  {\bibfnamefont {P.}~\bibnamefont {Coleman}}, \ and\ \bibinfo {author}
  {\bibfnamefont {S.}~\bibnamefont {Paschen}},\ }\href {\doibase 10.1103/PhysRevLett.104.186402} {\bibfield  {journal} {\bibinfo  {journal}
  {Phys. Rev. Lett.}\ }\textbf {\bibinfo {volume} {104}},\ \bibinfo {pages}
  {186402} (\bibinfo {year} {2010})}\BibitemShut {NoStop}%
\bibitem [{\citenamefont {Falkowski}\ and\ \citenamefont
  {Strydom}(2014)}]{Falkowski2014}%
  \BibitemOpen
  \bibfield  {author} {\bibinfo {author} {\bibfnamefont {M.}~\bibnamefont
  {Falkowski}}\ and\ \bibinfo {author} {\bibfnamefont {A.~M.}\ \bibnamefont
  {Strydom}},\ }\href {\doibase 10.1007/s10909-013-0907-5} {\bibfield
  {journal} {\bibinfo  {journal} {Journal of Low Temperature Physics}\ }\textbf
  {\bibinfo {volume} {175}},\ \bibinfo {pages} {498} (\bibinfo {year}
  {2014})}\BibitemShut {NoStop}%
\bibitem [{\citenamefont {Custers}\ \emph {et~al.}(2012)\citenamefont
  {Custers}, \citenamefont {Lorenzer}, \citenamefont {M{\"{u}}ller},
  \citenamefont {Prokofiev}, \citenamefont {Sidorenko}, \citenamefont
  {Winkler}, \citenamefont {Strydom}, \citenamefont {Shimura}, \citenamefont
  {Sakakibara}, \citenamefont {Yu}, \citenamefont {Si},\ and\ \citenamefont
  {Paschen}}]{Custers2012}%
  \BibitemOpen
  \bibfield  {author} {\bibinfo {author} {\bibfnamefont {J.}~\bibnamefont
  {Custers}}, \bibinfo {author} {\bibfnamefont {K.~A.}\ \bibnamefont
  {Lorenzer}}, \bibinfo {author} {\bibfnamefont {M.}~\bibnamefont
  {M{\"{u}}ller}}, \bibinfo {author} {\bibfnamefont {A.}~\bibnamefont
  {Prokofiev}}, \bibinfo {author} {\bibfnamefont {A.}~\bibnamefont
  {Sidorenko}}, \bibinfo {author} {\bibfnamefont {H.}~\bibnamefont {Winkler}},
  \bibinfo {author} {\bibfnamefont {A.~M.}\ \bibnamefont {Strydom}}, \bibinfo
  {author} {\bibfnamefont {Y.}~\bibnamefont {Shimura}}, \bibinfo {author}
  {\bibfnamefont {T.}~\bibnamefont {Sakakibara}}, \bibinfo {author}
  {\bibfnamefont {R.}~\bibnamefont {Yu}}, \bibinfo {author} {\bibfnamefont
  {Q.}~\bibnamefont {Si}}, \ and\ \bibinfo {author} {\bibfnamefont
  {S.}~\bibnamefont {Paschen}},\ }\href {\doibase 10.1038/nmat3214} {\bibfield
  {journal} {\bibinfo  {journal} {Nature Materials}\ }\textbf {\bibinfo
  {volume} {11}},\ \bibinfo {pages} {189} (\bibinfo {year} {2012})}\BibitemShut
  {NoStop}%
\bibitem [{\citenamefont {Cameron}\ \emph {et~al.}(2016)\citenamefont
  {Cameron}, \citenamefont {Friemel},\ and\ \citenamefont
  {Inosov}}]{Cameron2016}%
  \BibitemOpen
  \bibfield  {author} {\bibinfo {author} {\bibfnamefont {A.~S.}\ \bibnamefont
  {Cameron}}, \bibinfo {author} {\bibfnamefont {G.}~\bibnamefont {Friemel}}, \
  and\ \bibinfo {author} {\bibfnamefont {D.~S.}\ \bibnamefont {Inosov}},\
  }\href {\doibase 10.1088/0034-4885/79/6/066502} {\bibfield  {journal}
  {\bibinfo  {journal} {Reports on Progress in Physics}\ }\textbf {\bibinfo
  {volume} {79}} (\bibinfo {year} {2016}),\
  10.1088/0034-4885/79/6/066502}\BibitemShut {NoStop}%
\bibitem [{\citenamefont {Rylands}\ \emph {et~al.}(2022)\citenamefont
  {Rylands}, \citenamefont {Parhizkar},\ and\ \citenamefont
  {Galitski}}]{Rylands2022}%
  \BibitemOpen
  \bibfield  {author} {\bibinfo {author} {\bibfnamefont {C.}~\bibnamefont
  {Rylands}}, \bibinfo {author} {\bibfnamefont {A.}~\bibnamefont {Parhizkar}},
  \ and\ \bibinfo {author} {\bibfnamefont {V.}~\bibnamefont {Galitski}},\
  }\href {\doibase 10.1103/PhysRevB.105.195108} {\bibfield  {journal} {\bibinfo
   {journal} {Phys. Rev. B}\ }\textbf {\bibinfo {volume} {105}},\ \bibinfo
  {pages} {195108} (\bibinfo {year} {2022})}\BibitemShut {NoStop}%
\bibitem [{\citenamefont {Stewart}(2001)}]{Stewart2001}%
  \BibitemOpen
  \bibfield  {author} {\bibinfo {author} {\bibfnamefont {G.~R.}\ \bibnamefont
  {Stewart}},\ }\href {\doibase 10.1103/RevModPhys.73.797} {\bibfield
  {journal} {\bibinfo  {journal} {Reviews of Modern Physics}\ }\textbf
  {\bibinfo {volume} {73}},\ \bibinfo {pages} {797} (\bibinfo {year}
  {2001})}\BibitemShut {NoStop}%
\bibitem [{\citenamefont {L{\"{o}}hneysen}\ \emph {et~al.}(1994)\citenamefont
  {L{\"{o}}hneysen}, \citenamefont {Pietrus}, \citenamefont {Portisch},
  \citenamefont {Schlager}, \citenamefont {Schr{\"{o}}der}, \citenamefont
  {Sieck},\ and\ \citenamefont {Trappmann}}]{Lohneysen1994}%
  \BibitemOpen
  \bibfield  {author} {\bibinfo {author} {\bibfnamefont {H.~v.}\ \bibnamefont
  {L{\"{o}}hneysen}}, \bibinfo {author} {\bibfnamefont {T.}~\bibnamefont
  {Pietrus}}, \bibinfo {author} {\bibfnamefont {G.}~\bibnamefont {Portisch}},
  \bibinfo {author} {\bibfnamefont {H.~G.}\ \bibnamefont {Schlager}}, \bibinfo
  {author} {\bibfnamefont {A.}~\bibnamefont {Schr{\"{o}}der}}, \bibinfo
  {author} {\bibfnamefont {M.}~\bibnamefont {Sieck}}, \ and\ \bibinfo {author}
  {\bibfnamefont {T.}~\bibnamefont {Trappmann}},\ }\href {\doibase 10.1103/PhysRevLett.72.3262} {\bibfield  {journal} {\bibinfo  {journal}
  {Physical Review Letters}\ }\textbf {\bibinfo {volume} {72}},\ \bibinfo
  {pages} {3262} (\bibinfo {year} {1994})}\BibitemShut {NoStop}%
\bibitem [{\citenamefont {Abrahams}\ and\ \citenamefont
  {W{\"{o}}lfle}(2012)}]{Abrahams2012}%
  \BibitemOpen
  \bibfield  {author} {\bibinfo {author} {\bibfnamefont {E.}~\bibnamefont
  {Abrahams}}\ and\ \bibinfo {author} {\bibfnamefont {P.}~\bibnamefont
  {W{\"{o}}lfle}},\ }\href {\doibase 10.1073/pnas.1200346109} {\bibfield
  {journal} {\bibinfo  {journal} {Proceedings of the National Academy of
  Sciences of the United States of America}\ }\textbf {\bibinfo {volume}
  {109}},\ \bibinfo {pages} {3238} (\bibinfo {year} {2012})}\BibitemShut
  {NoStop}%
\bibitem [{\citenamefont {Irkhin}(2016)}]{Irkhin2016}%
  \BibitemOpen
  \bibfield  {author} {\bibinfo {author} {\bibfnamefont {V.~Y.}\ \bibnamefont
  {Irkhin}},\ }\href {\doibase 10.1140/epjb/e2016-60976-x} {\bibfield
  {journal} {\bibinfo  {journal} {The European Physical Journal B}\ }\textbf
  {\bibinfo {volume} {89}},\ \bibinfo {pages} {117} (\bibinfo {year}
  {2016})}\BibitemShut {NoStop}%
\bibitem [{\citenamefont {Irkhin}(2017)}]{Irkhin2017}%
  \BibitemOpen
  \bibfield  {author} {\bibinfo {author} {\bibfnamefont {V.~Y.}\ \bibnamefont
  {Irkhin}},\ }\href {\doibase 10.3367/ufne.2016.11.037961} {\bibfield
  {journal} {\bibinfo  {journal} {Physics-Uspekhi}\ }\textbf {\bibinfo {volume}
  {60}},\ \bibinfo {pages} {747} (\bibinfo {year} {2017})}\BibitemShut
  {NoStop}%
\bibitem [{\citenamefont {Iizuka}\ \emph {et~al.}()\citenamefont {Iizuka},
  \citenamefont {Yamada}, \citenamefont {Hanzawa},\ and\ \citenamefont
  {Ōno}}]{Iizuka2020}%
  \BibitemOpen
  \bibfield  {author} {\bibinfo {author} {\bibfnamefont {Y.}~\bibnamefont
  {Iizuka}}, \bibinfo {author} {\bibfnamefont {T.}~\bibnamefont {Yamada}},
  \bibinfo {author} {\bibfnamefont {K.}~\bibnamefont {Hanzawa}}, \ and\
  \bibinfo {author} {\bibfnamefont {Y.}~\bibnamefont {Ōno}},\ }\bibfield
  {booktitle} {\emph {\bibinfo {booktitle} {Proceedings of the International
  Conference on Strongly Correlated Electron Systems (SCES2019)}},\ }\href
  {\doibase 10.7566/JPSCP.30.011152} {\ 10.7566/JPSCP.30.011152}\BibitemShut
  {NoStop}%
\bibitem [{\citenamefont {Lai}\ \emph {et~al.}(2018)\citenamefont {Lai},
  \citenamefont {Nica}, \citenamefont {Hu}, \citenamefont {Gong}, \citenamefont
  {Paschen},\ and\ \citenamefont {Si}}]{Lai2018}%
  \BibitemOpen
  \bibfield  {author} {\bibinfo {author} {\bibfnamefont {H.-H.}\ \bibnamefont
  {Lai}}, \bibinfo {author} {\bibfnamefont {E.~M.}\ \bibnamefont {Nica}},
  \bibinfo {author} {\bibfnamefont {W.-J.}\ \bibnamefont {Hu}}, \bibinfo
  {author} {\bibfnamefont {S.-S.}\ \bibnamefont {Gong}}, \bibinfo {author}
  {\bibfnamefont {S.}~\bibnamefont {Paschen}}, \ and\ \bibinfo {author}
  {\bibfnamefont {Q.}~\bibnamefont {Si}},\ }\href {\doibase 10.48550/ARXIV.1807.09258} {\enquote {\bibinfo {title} {Kondo destruction and
  multipolar order-- implications for heavy fermion quantum criticality},}\ }
  (\bibinfo {year} {2018})\BibitemShut {NoStop}%
\bibitem [{\citenamefont {Kuzmenko}\ \emph {et~al.}(2018)\citenamefont
  {Kuzmenko}, \citenamefont {Kuzmenko}, \citenamefont {Avishai},\ and\
  \citenamefont {Jo}}]{Kuzmenko2018}%
  \BibitemOpen
  \bibfield  {author} {\bibinfo {author} {\bibfnamefont {I.}~\bibnamefont
  {Kuzmenko}}, \bibinfo {author} {\bibfnamefont {T.}~\bibnamefont {Kuzmenko}},
  \bibinfo {author} {\bibfnamefont {Y.}~\bibnamefont {Avishai}}, \ and\
  \bibinfo {author} {\bibfnamefont {G.-B.}\ \bibnamefont {Jo}},\ }\href
  {\doibase 10.1103/PhysRevB.97.075124} {\bibfield  {journal} {\bibinfo
  {journal} {Phys. Rev. B}\ }\textbf {\bibinfo {volume} {97}},\ \bibinfo
  {pages} {075124} (\bibinfo {year} {2018})}\BibitemShut {NoStop}%
\bibitem [{\citenamefont {Onimaru}\ \emph {et~al.}(2011)\citenamefont
  {Onimaru}, \citenamefont {Matsumoto}, \citenamefont {Inoue}, \citenamefont
  {Umeo}, \citenamefont {Sakakibara}, \citenamefont {Karaki}, \citenamefont
  {Kubota},\ and\ \citenamefont {Takabatake}}]{Onimaru2011a}%
  \BibitemOpen
  \bibfield  {author} {\bibinfo {author} {\bibfnamefont {T.}~\bibnamefont
  {Onimaru}}, \bibinfo {author} {\bibfnamefont {K.~T.}\ \bibnamefont
  {Matsumoto}}, \bibinfo {author} {\bibfnamefont {Y.~F.}\ \bibnamefont
  {Inoue}}, \bibinfo {author} {\bibfnamefont {K.}~\bibnamefont {Umeo}},
  \bibinfo {author} {\bibfnamefont {T.}~\bibnamefont {Sakakibara}}, \bibinfo
  {author} {\bibfnamefont {Y.}~\bibnamefont {Karaki}}, \bibinfo {author}
  {\bibfnamefont {M.}~\bibnamefont {Kubota}}, \ and\ \bibinfo {author}
  {\bibfnamefont {T.}~\bibnamefont {Takabatake}},\ }\href {\doibase 10.1103/PhysRevLett.106.177001} {\bibfield  {journal} {\bibinfo  {journal}
  {Phys. Rev. Lett.}\ }\textbf {\bibinfo {volume} {106}},\ \bibinfo {pages}
  {177001} (\bibinfo {year} {2011})}\BibitemShut {NoStop}%
\bibitem [{\citenamefont {Onimaru}\ and\ \citenamefont
  {Kusunose}(2016)}]{Onimaru2016b}%
  \BibitemOpen
  \bibfield  {author} {\bibinfo {author} {\bibfnamefont {T.}~\bibnamefont
  {Onimaru}}\ and\ \bibinfo {author} {\bibfnamefont {H.}~\bibnamefont
  {Kusunose}},\ }\href {\doibase 10.7566/JPSJ.85.082002} {\bibfield  {journal}
  {\bibinfo  {journal} {Journal of the Physical Society of Japan}\ }\textbf
  {\bibinfo {volume} {85}},\ \bibinfo {pages} {082002} (\bibinfo {year}
  {2016})}\BibitemShut {NoStop}%
\bibitem [{\citenamefont {Onimaru}\ \emph {et~al.}(2016)\citenamefont
  {Onimaru}, \citenamefont {Izawa}, \citenamefont {Matsumoto}, \citenamefont
  {Yoshida}, \citenamefont {Machida}, \citenamefont {Ikeura}, \citenamefont
  {Wakiya}, \citenamefont {Umeo}, \citenamefont {Kittaka}, \citenamefont
  {Araki}, \citenamefont {Sakakibara},\ and\ \citenamefont
  {Takabatake}}]{Onimaru2016c}%
  \BibitemOpen
  \bibfield  {author} {\bibinfo {author} {\bibfnamefont {T.}~\bibnamefont
  {Onimaru}}, \bibinfo {author} {\bibfnamefont {K.}~\bibnamefont {Izawa}},
  \bibinfo {author} {\bibfnamefont {K.~T.}\ \bibnamefont {Matsumoto}}, \bibinfo
  {author} {\bibfnamefont {T.}~\bibnamefont {Yoshida}}, \bibinfo {author}
  {\bibfnamefont {Y.}~\bibnamefont {Machida}}, \bibinfo {author} {\bibfnamefont
  {T.}~\bibnamefont {Ikeura}}, \bibinfo {author} {\bibfnamefont
  {K.}~\bibnamefont {Wakiya}}, \bibinfo {author} {\bibfnamefont
  {K.}~\bibnamefont {Umeo}}, \bibinfo {author} {\bibfnamefont {S.}~\bibnamefont
  {Kittaka}}, \bibinfo {author} {\bibfnamefont {K.}~\bibnamefont {Araki}},
  \bibinfo {author} {\bibfnamefont {T.}~\bibnamefont {Sakakibara}}, \ and\
  \bibinfo {author} {\bibfnamefont {T.}~\bibnamefont {Takabatake}},\ }\href
  {\doibase 10.1103/PhysRevB.94.075134} {\bibfield  {journal} {\bibinfo
  {journal} {Phys. Rev. B}\ }\textbf {\bibinfo {volume} {94}},\ \bibinfo
  {pages} {075134} (\bibinfo {year} {2016})}\BibitemShut {NoStop}%
\bibitem [{\citenamefont {Freyer}\ \emph {et~al.}(2018)\citenamefont {Freyer},
  \citenamefont {Attig}, \citenamefont {Lee}, \citenamefont {Paramekanti},
  \citenamefont {Trebst},\ and\ \citenamefont {Kim}}]{Freyer2018a}%
  \BibitemOpen
  \bibfield  {author} {\bibinfo {author} {\bibfnamefont {F.}~\bibnamefont
  {Freyer}}, \bibinfo {author} {\bibfnamefont {J.}~\bibnamefont {Attig}},
  \bibinfo {author} {\bibfnamefont {S.}~\bibnamefont {Lee}}, \bibinfo {author}
  {\bibfnamefont {A.}~\bibnamefont {Paramekanti}}, \bibinfo {author}
  {\bibfnamefont {S.}~\bibnamefont {Trebst}}, \ and\ \bibinfo {author}
  {\bibfnamefont {Y.~B.}\ \bibnamefont {Kim}},\ }\href {\doibase 10.1103/PhysRevB.97.115111} {\bibfield  {journal} {\bibinfo  {journal}
  {Physical Review B}\ }\textbf {\bibinfo {volume} {97}},\ \bibinfo {pages}
  {115111} (\bibinfo {year} {2018})}\BibitemShut {NoStop}%
\bibitem [{\citenamefont {Sato}\ \emph {et~al.}(2012)\citenamefont {Sato},
  \citenamefont {Ibuka}, \citenamefont {Nambu}, \citenamefont {Yamazaki},
  \citenamefont {Hong}, \citenamefont {Sakai},\ and\ \citenamefont
  {Nakatsuji}}]{Sato2012a}%
  \BibitemOpen
  \bibfield  {author} {\bibinfo {author} {\bibfnamefont {T.~J.}\ \bibnamefont
  {Sato}}, \bibinfo {author} {\bibfnamefont {S.}~\bibnamefont {Ibuka}},
  \bibinfo {author} {\bibfnamefont {Y.}~\bibnamefont {Nambu}}, \bibinfo
  {author} {\bibfnamefont {T.}~\bibnamefont {Yamazaki}}, \bibinfo {author}
  {\bibfnamefont {T.}~\bibnamefont {Hong}}, \bibinfo {author} {\bibfnamefont
  {A.}~\bibnamefont {Sakai}}, \ and\ \bibinfo {author} {\bibfnamefont
  {S.}~\bibnamefont {Nakatsuji}},\ }\href {\doibase 10.1103/PhysRevB.86.184419}
  {\bibfield  {journal} {\bibinfo  {journal} {Phys. Rev. B}\ }\textbf {\bibinfo
  {volume} {86}},\ \bibinfo {pages} {184419} (\bibinfo {year}
  {2012})}\BibitemShut {NoStop}%
\bibitem [{\citenamefont {Araki}\ \emph {et~al.}()\citenamefont {Araki},
  \citenamefont {Shimura}, \citenamefont {Kase}, \citenamefont {Sakakibara},
  \citenamefont {Sakai},\ and\ \citenamefont {Nakatsuji}}]{Araki2014}%
  \BibitemOpen
  \bibfield  {author} {\bibinfo {author} {\bibfnamefont {K.}~\bibnamefont
  {Araki}}, \bibinfo {author} {\bibfnamefont {Y.}~\bibnamefont {Shimura}},
  \bibinfo {author} {\bibfnamefont {N.}~\bibnamefont {Kase}}, \bibinfo {author}
  {\bibfnamefont {T.}~\bibnamefont {Sakakibara}}, \bibinfo {author}
  {\bibfnamefont {A.}~\bibnamefont {Sakai}}, \ and\ \bibinfo {author}
  {\bibfnamefont {S.}~\bibnamefont {Nakatsuji}},\ }\bibfield  {booktitle}
  {\emph {\bibinfo {booktitle} {Proceedings of the International Conference on
  Strongly Correlated Electron Systems (SCES2013)}},\ }\href {\doibase 10.7566/JPSCP.3.011093} {\ 10.7566/JPSCP.3.011093}\BibitemShut {NoStop}%
\bibitem [{\citenamefont {Matsushita}\ \emph {et~al.}(2011)\citenamefont
  {Matsushita}, \citenamefont {Sakaguchi}, \citenamefont {Taga}, \citenamefont
  {Ohya}, \citenamefont {Yoshiuchi}, \citenamefont {Ota}, \citenamefont
  {Hirose}, \citenamefont {Enoki}, \citenamefont {Honda}, \citenamefont
  {Sugiyama}, \citenamefont {Hagiwara}, \citenamefont {Kindo}, \citenamefont
  {Tanaka}, \citenamefont {Kubo}, \citenamefont {Takeuchi}, \citenamefont
  {Settai},\ and\ \citenamefont {Ōnuki}}]{Matsushita2011}%
  \BibitemOpen
  \bibfield  {author} {\bibinfo {author} {\bibfnamefont {M.}~\bibnamefont
  {Matsushita}}, \bibinfo {author} {\bibfnamefont {J.}~\bibnamefont
  {Sakaguchi}}, \bibinfo {author} {\bibfnamefont {Y.}~\bibnamefont {Taga}},
  \bibinfo {author} {\bibfnamefont {M.}~\bibnamefont {Ohya}}, \bibinfo {author}
  {\bibfnamefont {S.}~\bibnamefont {Yoshiuchi}}, \bibinfo {author}
  {\bibfnamefont {H.}~\bibnamefont {Ota}}, \bibinfo {author} {\bibfnamefont
  {Y.}~\bibnamefont {Hirose}}, \bibinfo {author} {\bibfnamefont
  {K.}~\bibnamefont {Enoki}}, \bibinfo {author} {\bibfnamefont
  {F.}~\bibnamefont {Honda}}, \bibinfo {author} {\bibfnamefont
  {K.}~\bibnamefont {Sugiyama}}, \bibinfo {author} {\bibfnamefont
  {M.}~\bibnamefont {Hagiwara}}, \bibinfo {author} {\bibfnamefont
  {K.}~\bibnamefont {Kindo}}, \bibinfo {author} {\bibfnamefont
  {T.}~\bibnamefont {Tanaka}}, \bibinfo {author} {\bibfnamefont
  {Y.}~\bibnamefont {Kubo}}, \bibinfo {author} {\bibfnamefont {T.}~\bibnamefont
  {Takeuchi}}, \bibinfo {author} {\bibfnamefont {R.}~\bibnamefont {Settai}}, \
  and\ \bibinfo {author} {\bibfnamefont {Y.}~\bibnamefont {Ōnuki}},\ }\href
  {\doibase 10.1143/JPSJ.80.074605} {\bibfield  {journal} {\bibinfo  {journal}
  {Journal of the Physical Society of Japan}\ }\textbf {\bibinfo {volume}
  {80}},\ \bibinfo {pages} {074605} (\bibinfo {year} {2011})}\BibitemShut
  {NoStop}%
\bibitem [{\citenamefont {Kumar}\ \emph {et~al.}(2016)\citenamefont {Kumar},
  \citenamefont {Nair}, \citenamefont {Christian}, \citenamefont
  {Thamizhavel},\ and\ \citenamefont {Strydom}}]{RameshKumar2016}%
  \BibitemOpen
  \bibfield  {author} {\bibinfo {author} {\bibfnamefont {K.~R.}\ \bibnamefont
  {Kumar}}, \bibinfo {author} {\bibfnamefont {H.~S.}\ \bibnamefont {Nair}},
  \bibinfo {author} {\bibfnamefont {R.}~\bibnamefont {Christian}}, \bibinfo
  {author} {\bibfnamefont {A.}~\bibnamefont {Thamizhavel}}, \ and\ \bibinfo
  {author} {\bibfnamefont {A.~M.}\ \bibnamefont {Strydom}},\ }\href {\doibase 10.1088/0953-8984/28/43/436002} {\bibfield  {journal} {\bibinfo  {journal}
  {Journal of Physics: Condensed Matter}\ }\textbf {\bibinfo {volume} {28}},\
  \bibinfo {pages} {436002} (\bibinfo {year} {2016})}\BibitemShut {NoStop}%
\bibitem [{\citenamefont {Sakai}\ and\ \citenamefont
  {Nakatsuji}(2011)}]{Sakai2011b}%
  \BibitemOpen
  \bibfield  {author} {\bibinfo {author} {\bibfnamefont {A.}~\bibnamefont
  {Sakai}}\ and\ \bibinfo {author} {\bibfnamefont {S.}~\bibnamefont
  {Nakatsuji}},\ }\href {\doibase 10.1143/JPSJ.80.063701} {\bibfield  {journal}
  {\bibinfo  {journal} {Journal of the Physical Society of Japan}\ }\textbf
  {\bibinfo {volume} {80}},\ \bibinfo {pages} {063701} (\bibinfo {year}
  {2011})}\BibitemShut {NoStop}%
\bibitem [{\citenamefont {W\"orl}\ \emph {et~al.}(2019)\citenamefont {W\"orl},
  \citenamefont {Onimaru}, \citenamefont {Tokiwa}, \citenamefont {Yamane},
  \citenamefont {Matsumoto}, \citenamefont {Takabatake},\ and\ \citenamefont
  {Gegenwart}}]{Worl2019a}%
  \BibitemOpen
  \bibfield  {author} {\bibinfo {author} {\bibfnamefont {A.}~\bibnamefont
  {W\"orl}}, \bibinfo {author} {\bibfnamefont {T.}~\bibnamefont {Onimaru}},
  \bibinfo {author} {\bibfnamefont {Y.}~\bibnamefont {Tokiwa}}, \bibinfo
  {author} {\bibfnamefont {Y.}~\bibnamefont {Yamane}}, \bibinfo {author}
  {\bibfnamefont {K.~T.}\ \bibnamefont {Matsumoto}}, \bibinfo {author}
  {\bibfnamefont {T.}~\bibnamefont {Takabatake}}, \ and\ \bibinfo {author}
  {\bibfnamefont {P.}~\bibnamefont {Gegenwart}},\ }\href {\doibase 10.1103/PhysRevB.99.081117} {\bibfield  {journal} {\bibinfo  {journal} {Phys.
  Rev. B}\ }\textbf {\bibinfo {volume} {99}},\ \bibinfo {pages} {081117}
  (\bibinfo {year} {2019})}\BibitemShut {NoStop}%
\bibitem [{\citenamefont {Matsubayashi}\ \emph {et~al.}(2012)\citenamefont
  {Matsubayashi}, \citenamefont {Tanaka}, \citenamefont {Sakai}, \citenamefont
  {Nakatsuji}, \citenamefont {Kubo},\ and\ \citenamefont
  {Uwatoko}}]{Matsubayashi2012c}%
  \BibitemOpen
  \bibfield  {author} {\bibinfo {author} {\bibfnamefont {K.}~\bibnamefont
  {Matsubayashi}}, \bibinfo {author} {\bibfnamefont {T.}~\bibnamefont
  {Tanaka}}, \bibinfo {author} {\bibfnamefont {A.}~\bibnamefont {Sakai}},
  \bibinfo {author} {\bibfnamefont {S.}~\bibnamefont {Nakatsuji}}, \bibinfo
  {author} {\bibfnamefont {Y.}~\bibnamefont {Kubo}}, \ and\ \bibinfo {author}
  {\bibfnamefont {Y.}~\bibnamefont {Uwatoko}},\ }\href {\doibase 10.1103/PhysRevLett.109.187004} {\bibfield  {journal} {\bibinfo  {journal}
  {Phys. Rev. Lett.}\ }\textbf {\bibinfo {volume} {109}},\ \bibinfo {pages}
  {187004} (\bibinfo {year} {2012})}\BibitemShut {NoStop}%
\bibitem [{\citenamefont {Sakai}\ \emph {et~al.}(2012)\citenamefont {Sakai},
  \citenamefont {Kuga},\ and\ \citenamefont {Nakatsuji}}]{Sakai2012a}%
  \BibitemOpen
  \bibfield  {author} {\bibinfo {author} {\bibfnamefont {A.}~\bibnamefont
  {Sakai}}, \bibinfo {author} {\bibfnamefont {K.}~\bibnamefont {Kuga}}, \ and\
  \bibinfo {author} {\bibfnamefont {S.}~\bibnamefont {Nakatsuji}},\ }\href
  {\doibase 10.1143/JPSJ.81.083702} {\bibfield  {journal} {\bibinfo  {journal}
  {Journal of the Physical Society of Japan}\ }\textbf {\bibinfo {volume}
  {81}},\ \bibinfo {pages} {083702} (\bibinfo {year} {2012})}\BibitemShut
  {NoStop}%
\bibitem [{\citenamefont {Tsujimoto}\ \emph {et~al.}(2014)\citenamefont
  {Tsujimoto}, \citenamefont {Matsumoto}, \citenamefont {Tomita}, \citenamefont
  {Sakai},\ and\ \citenamefont {Nakatsuji}}]{Tsujimoto2014a}%
  \BibitemOpen
  \bibfield  {author} {\bibinfo {author} {\bibfnamefont {M.}~\bibnamefont
  {Tsujimoto}}, \bibinfo {author} {\bibfnamefont {Y.}~\bibnamefont
  {Matsumoto}}, \bibinfo {author} {\bibfnamefont {T.}~\bibnamefont {Tomita}},
  \bibinfo {author} {\bibfnamefont {A.}~\bibnamefont {Sakai}}, \ and\ \bibinfo
  {author} {\bibfnamefont {S.}~\bibnamefont {Nakatsuji}},\ }\href {\doibase 10.1103/PhysRevLett.113.267001} {\bibfield  {journal} {\bibinfo  {journal}
  {Phys. Rev. Lett.}\ }\textbf {\bibinfo {volume} {113}},\ \bibinfo {pages}
  {267001} (\bibinfo {year} {2014})}\BibitemShut {NoStop}%
\bibitem [{\citenamefont {Fu}\ \emph {et~al.}(2020)\citenamefont {Fu},
  \citenamefont {Sakai}, \citenamefont {Sogabe}, \citenamefont {Tsujimoto},
  \citenamefont {Matsumoto},\ and\ \citenamefont {Nakatsuji}}]{Fu2020a}%
  \BibitemOpen
  \bibfield  {author} {\bibinfo {author} {\bibfnamefont {M.}~\bibnamefont
  {Fu}}, \bibinfo {author} {\bibfnamefont {A.}~\bibnamefont {Sakai}}, \bibinfo
  {author} {\bibfnamefont {N.}~\bibnamefont {Sogabe}}, \bibinfo {author}
  {\bibfnamefont {M.}~\bibnamefont {Tsujimoto}}, \bibinfo {author}
  {\bibfnamefont {Y.}~\bibnamefont {Matsumoto}}, \ and\ \bibinfo {author}
  {\bibfnamefont {S.}~\bibnamefont {Nakatsuji}},\ }\href {\doibase 10.7566/JPSJ.89.013704} {\bibfield  {journal} {\bibinfo  {journal} {Journal
  of the Physical Society of Japan}\ }\textbf {\bibinfo {volume} {89}},\
  \bibinfo {pages} {013704} (\bibinfo {year} {2020})}\BibitemShut {NoStop}%
\bibitem [{\citenamefont {Onimaru}\ \emph {et~al.}(2012)\citenamefont
  {Onimaru}, \citenamefont {Nagasawa}, \citenamefont {Matsumoto}, \citenamefont
  {Wakiya}, \citenamefont {Umeo}, \citenamefont {Kittaka}, \citenamefont
  {Sakakibara}, \citenamefont {Matsushita},\ and\ \citenamefont
  {Takabatake}}]{Onimaru2012}%
  \BibitemOpen
  \bibfield  {author} {\bibinfo {author} {\bibfnamefont {T.}~\bibnamefont
  {Onimaru}}, \bibinfo {author} {\bibfnamefont {N.}~\bibnamefont {Nagasawa}},
  \bibinfo {author} {\bibfnamefont {K.~T.}\ \bibnamefont {Matsumoto}}, \bibinfo
  {author} {\bibfnamefont {K.}~\bibnamefont {Wakiya}}, \bibinfo {author}
  {\bibfnamefont {K.}~\bibnamefont {Umeo}}, \bibinfo {author} {\bibfnamefont
  {S.}~\bibnamefont {Kittaka}}, \bibinfo {author} {\bibfnamefont
  {T.}~\bibnamefont {Sakakibara}}, \bibinfo {author} {\bibfnamefont
  {Y.}~\bibnamefont {Matsushita}}, \ and\ \bibinfo {author} {\bibfnamefont
  {T.}~\bibnamefont {Takabatake}},\ }\href {\doibase 10.1103/PhysRevB.86.184426} {\bibfield  {journal} {\bibinfo  {journal} {Phys.
  Rev. B}\ }\textbf {\bibinfo {volume} {86}},\ \bibinfo {pages} {184426}
  (\bibinfo {year} {2012})}\BibitemShut {NoStop}%
\bibitem [{\citenamefont {Matsumoto}\ \emph {et~al.}(2015)\citenamefont
  {Matsumoto}, \citenamefont {Onimaru}, \citenamefont {Wakiya}, \citenamefont
  {Umeo},\ and\ \citenamefont {Takabatake}}]{Matsumoto2015}%
  \BibitemOpen
  \bibfield  {author} {\bibinfo {author} {\bibfnamefont {K.~T.}\ \bibnamefont
  {Matsumoto}}, \bibinfo {author} {\bibfnamefont {T.}~\bibnamefont {Onimaru}},
  \bibinfo {author} {\bibfnamefont {K.}~\bibnamefont {Wakiya}}, \bibinfo
  {author} {\bibfnamefont {K.}~\bibnamefont {Umeo}}, \ and\ \bibinfo {author}
  {\bibfnamefont {T.}~\bibnamefont {Takabatake}},\ }\href {\doibase 10.7566/JPSJ.84.063703} {\bibfield  {journal} {\bibinfo  {journal} {Journal
  of the Physical Society of Japan}\ }\textbf {\bibinfo {volume} {84}},\
  \bibinfo {pages} {063703} (\bibinfo {year} {2015})}\BibitemShut {NoStop}%
\bibitem [{\citenamefont {Onimaru}\ \emph {et~al.}(2010)\citenamefont
  {Onimaru}, \citenamefont {T.~Matsumoto}, \citenamefont {F.~Inoue},
  \citenamefont {Umeo}, \citenamefont {Saiga}, \citenamefont {Matsushita},
  \citenamefont {Tamura}, \citenamefont {Nishimoto}, \citenamefont {Ishii},
  \citenamefont {Suzuki},\ and\ \citenamefont {Takabatake}}]{Onimaru2010}%
  \BibitemOpen
  \bibfield  {author} {\bibinfo {author} {\bibfnamefont {T.}~\bibnamefont
  {Onimaru}}, \bibinfo {author} {\bibfnamefont {K.}~\bibnamefont
  {T.~Matsumoto}}, \bibinfo {author} {\bibfnamefont {Y.}~\bibnamefont
  {F.~Inoue}}, \bibinfo {author} {\bibfnamefont {K.}~\bibnamefont {Umeo}},
  \bibinfo {author} {\bibfnamefont {Y.}~\bibnamefont {Saiga}}, \bibinfo
  {author} {\bibfnamefont {Y.}~\bibnamefont {Matsushita}}, \bibinfo {author}
  {\bibfnamefont {R.}~\bibnamefont {Tamura}}, \bibinfo {author} {\bibfnamefont
  {K.}~\bibnamefont {Nishimoto}}, \bibinfo {author} {\bibfnamefont
  {I.}~\bibnamefont {Ishii}}, \bibinfo {author} {\bibfnamefont
  {T.}~\bibnamefont {Suzuki}}, \ and\ \bibinfo {author} {\bibfnamefont
  {T.}~\bibnamefont {Takabatake}},\ }\href {\doibase 10.1143/JPSJ.79.033704}
  {\bibfield  {journal} {\bibinfo  {journal} {Journal of the Physical Society
  of Japan}\ }\textbf {\bibinfo {volume} {79}},\ \bibinfo {pages} {033704}
  (\bibinfo {year} {2010})}\BibitemShut {NoStop}%
\bibitem [{\citenamefont {Shimura}\ \emph {et~al.}(2015)\citenamefont
  {Shimura}, \citenamefont {Tsujimoto}, \citenamefont {Zeng}, \citenamefont
  {Balicas}, \citenamefont {Sakai},\ and\ \citenamefont
  {Nakatsuji}}]{Shimura2015a}%
  \BibitemOpen
  \bibfield  {author} {\bibinfo {author} {\bibfnamefont {Y.}~\bibnamefont
  {Shimura}}, \bibinfo {author} {\bibfnamefont {M.}~\bibnamefont {Tsujimoto}},
  \bibinfo {author} {\bibfnamefont {B.}~\bibnamefont {Zeng}}, \bibinfo {author}
  {\bibfnamefont {L.}~\bibnamefont {Balicas}}, \bibinfo {author} {\bibfnamefont
  {A.}~\bibnamefont {Sakai}}, \ and\ \bibinfo {author} {\bibfnamefont
  {S.}~\bibnamefont {Nakatsuji}},\ }\href {\doibase 10.1103/PhysRevB.91.241102}
  {\bibfield  {journal} {\bibinfo  {journal} {Phys. Rev. B}\ }\textbf {\bibinfo
  {volume} {91}},\ \bibinfo {pages} {241102} (\bibinfo {year}
  {2015})}\BibitemShut {NoStop}%
\bibitem [{\citenamefont {Nagashima}\ \emph {et~al.}(2014)\citenamefont
  {Nagashima}, \citenamefont {Nishiwaki}, \citenamefont {Otani}, \citenamefont
  {Sakoda}, \citenamefont {Matsuoka}, \citenamefont {Harima},\ and\
  \citenamefont {Sugawara}}]{Nagashima2014a}%
  \BibitemOpen
  \bibfield  {author} {\bibinfo {author} {\bibfnamefont {S.}~\bibnamefont
  {Nagashima}}, \bibinfo {author} {\bibfnamefont {T.}~\bibnamefont
  {Nishiwaki}}, \bibinfo {author} {\bibfnamefont {A.}~\bibnamefont {Otani}},
  \bibinfo {author} {\bibfnamefont {M.}~\bibnamefont {Sakoda}}, \bibinfo
  {author} {\bibfnamefont {E.}~\bibnamefont {Matsuoka}}, \bibinfo {author}
  {\bibfnamefont {H.}~\bibnamefont {Harima}}, \ and\ \bibinfo {author}
  {\bibfnamefont {H.}~\bibnamefont {Sugawara}},\ }in\ \href {\doibase 10.7566/JPSCP.3.011019} {\emph {\bibinfo {booktitle} {Proceedings of the
  International Conference on Strongly Correlated Electron Systems
  (SCES2013)}}},\ Vol.\ \bibinfo {volume} {011019}\ (\bibinfo  {publisher}
  {Journal of the Physical Society of Japan},\ \bibinfo {year} {2014})\ pp.\
  \bibinfo {pages} {2--7}\BibitemShut {NoStop}%
\bibitem [{\citenamefont {Schultz}\ \emph
  {et~al.}(2021{\natexlab{a}})\citenamefont {Schultz}, \citenamefont {Patri},\
  and\ \citenamefont {Kim}}]{Schultz2021b}%
  \BibitemOpen
  \bibfield  {author} {\bibinfo {author} {\bibfnamefont {D.~J.}\ \bibnamefont
  {Schultz}}, \bibinfo {author} {\bibfnamefont {A.~S.}\ \bibnamefont {Patri}},
  \ and\ \bibinfo {author} {\bibfnamefont {Y.~B.}\ \bibnamefont {Kim}},\ }\href
  {\doibase 10.1103/PhysRevResearch.3.013189} {\bibfield  {journal} {\bibinfo
  {journal} {Physical Review Research}\ }\textbf {\bibinfo {volume} {3}},\
  \bibinfo {pages} {013189} (\bibinfo {year} {2021}{\natexlab{a}})}\BibitemShut
  {NoStop}%
\bibitem [{\citenamefont {Patri}\ and\ \citenamefont {Kim}(2020)}]{Patri2020e}%
  \BibitemOpen
  \bibfield  {author} {\bibinfo {author} {\bibfnamefont {A.~S.}\ \bibnamefont
  {Patri}}\ and\ \bibinfo {author} {\bibfnamefont {Y.~B.}\ \bibnamefont
  {Kim}},\ }\href {\doibase 10.1103/PhysRevX.10.041021} {\bibfield  {journal}
  {\bibinfo  {journal} {Phys. Rev. X}\ }\textbf {\bibinfo {volume} {10}},\
  \bibinfo {pages} {041021} (\bibinfo {year} {2020})}\BibitemShut {NoStop}%
\bibitem [{\citenamefont {Patri}\ and\ \citenamefont {Kim}(2022)}]{Patri2022}%
  \BibitemOpen
  \bibfield  {author} {\bibinfo {author} {\bibfnamefont {A.~S.}\ \bibnamefont
  {Patri}}\ and\ \bibinfo {author} {\bibfnamefont {Y.~B.}\ \bibnamefont
  {Kim}},\ }\href {\doibase 10.21468/SciPostPhys.12.2.057} {\bibfield
  {journal} {\bibinfo  {journal} {SciPost Phys.}\ }\textbf {\bibinfo {volume}
  {12}},\ \bibinfo {pages} {057} (\bibinfo {year} {2022})}\BibitemShut
  {NoStop}%
\bibitem [{\citenamefont {Patri}\ \emph {et~al.}(2019)\citenamefont {Patri},
  \citenamefont {Sakai}, \citenamefont {Lee}, \citenamefont {Paramekanti},
  \citenamefont {Nakatsuji},\ and\ \citenamefont {Kim}}]{Patri2019d}%
  \BibitemOpen
  \bibfield  {author} {\bibinfo {author} {\bibfnamefont {A.~S.}\ \bibnamefont
  {Patri}}, \bibinfo {author} {\bibfnamefont {A.}~\bibnamefont {Sakai}},
  \bibinfo {author} {\bibfnamefont {S.}~\bibnamefont {Lee}}, \bibinfo {author}
  {\bibfnamefont {A.}~\bibnamefont {Paramekanti}}, \bibinfo {author}
  {\bibfnamefont {S.}~\bibnamefont {Nakatsuji}}, \ and\ \bibinfo {author}
  {\bibfnamefont {Y.~B.}\ \bibnamefont {Kim}},\ }\href {\doibase 10.1038/s41467-019-11913-3} {\bibfield  {journal} {\bibinfo  {journal}
  {Nature Communications}\ }\textbf {\bibinfo {volume} {10}},\ \bibinfo {pages}
  {4092} (\bibinfo {year} {2019})}\BibitemShut {NoStop}%
\bibitem [{\citenamefont {Tsuruta}\ and\ \citenamefont
  {Miyake}(2015)}]{Tsuruta2015}%
  \BibitemOpen
  \bibfield  {author} {\bibinfo {author} {\bibfnamefont {A.}~\bibnamefont
  {Tsuruta}}\ and\ \bibinfo {author} {\bibfnamefont {K.}~\bibnamefont
  {Miyake}},\ }\href {\doibase 10.7566/JPSJ.84.114714} {\bibfield  {journal}
  {\bibinfo  {journal} {Journal of the Physical Society of Japan}\ }\textbf
  {\bibinfo {volume} {84}},\ \bibinfo {pages} {114714} (\bibinfo {year}
  {2015})}\BibitemShut {NoStop}%
\bibitem [{\citenamefont {Tsuruta}\ \emph {et~al.}(1999)\citenamefont
  {Tsuruta}, \citenamefont {Kobayashi}, \citenamefont {{\"{O}}no},
  \citenamefont {Matsuura},\ and\ \citenamefont {Kuroda}}]{Tsuruta1999}%
  \BibitemOpen
  \bibfield  {author} {\bibinfo {author} {\bibfnamefont {A.}~\bibnamefont
  {Tsuruta}}, \bibinfo {author} {\bibfnamefont {A.}~\bibnamefont {Kobayashi}},
  \bibinfo {author} {\bibfnamefont {Y.}~\bibnamefont {{\"{O}}no}}, \bibinfo
  {author} {\bibfnamefont {T.}~\bibnamefont {Matsuura}}, \ and\ \bibinfo
  {author} {\bibfnamefont {Y.}~\bibnamefont {Kuroda}},\ }\href {\doibase 10.1143/JPSJ.68.2491} {\bibfield  {journal} {\bibinfo  {journal} {Journal of
  the Physical Society of Japan}\ }\textbf {\bibinfo {volume} {68}},\ \bibinfo
  {pages} {2491} (\bibinfo {year} {1999})}\BibitemShut {NoStop}%
\bibitem [{\citenamefont {Tsuruta}\ \emph
  {et~al.}(2000{\natexlab{a}})\citenamefont {Tsuruta}, \citenamefont
  {Kobayashi}, \citenamefont {Matsuura},\ and\ \citenamefont
  {Kuroda}}]{Tsuruta2000a}%
  \BibitemOpen
  \bibfield  {author} {\bibinfo {author} {\bibfnamefont {A.}~\bibnamefont
  {Tsuruta}}, \bibinfo {author} {\bibfnamefont {A.}~\bibnamefont {Kobayashi}},
  \bibinfo {author} {\bibfnamefont {T.}~\bibnamefont {Matsuura}}, \ and\
  \bibinfo {author} {\bibfnamefont {Y.}~\bibnamefont {Kuroda}},\ }\href
  {\doibase 10.1143/JPSJ.69.3342} {\bibfield  {journal} {\bibinfo  {journal}
  {Journal of the Physical Society of Japan}\ }\textbf {\bibinfo {volume}
  {69}},\ \bibinfo {pages} {3342} (\bibinfo {year}
  {2000}{\natexlab{a}})}\BibitemShut {NoStop}%
\bibitem [{\citenamefont {Tsuruta}\ and\ \citenamefont
  {Miyake}(2022)}]{Tsuruta2022}%
  \BibitemOpen
  \bibfield  {author} {\bibinfo {author} {\bibfnamefont {A.}~\bibnamefont
  {Tsuruta}}\ and\ \bibinfo {author} {\bibfnamefont {K.}~\bibnamefont
  {Miyake}},\ }\href {\doibase 10.1088/2399-6528/ac42dd} {\bibfield  {journal}
  {\bibinfo  {journal} {Journal of Physics Communications}\ }\textbf {\bibinfo
  {volume} {6}},\ \bibinfo {pages} {015006} (\bibinfo {year}
  {2022})}\BibitemShut {NoStop}%
\bibitem [{\citenamefont {Patri}\ \emph {et~al.}(2020)\citenamefont {Patri},
  \citenamefont {Khait},\ and\ \citenamefont {Kim}}]{Patri2020d}%
  \BibitemOpen
  \bibfield  {author} {\bibinfo {author} {\bibfnamefont {A.~S.}\ \bibnamefont
  {Patri}}, \bibinfo {author} {\bibfnamefont {I.}~\bibnamefont {Khait}}, \ and\
  \bibinfo {author} {\bibfnamefont {Y.~B.}\ \bibnamefont {Kim}},\ }\href
  {\doibase 10.1103/PhysRevResearch.2.013257} {\bibfield  {journal} {\bibinfo
  {journal} {Physical Review Research}\ }\textbf {\bibinfo {volume} {2}},\
  \bibinfo {pages} {013257} (\bibinfo {year} {2020})}\BibitemShut {NoStop}%
\bibitem [{\citenamefont {Lee}\ \emph {et~al.}(2018)\citenamefont {Lee},
  \citenamefont {Trebst}, \citenamefont {Kim},\ and\ \citenamefont
  {Paramekanti}}]{Lee2018e}%
  \BibitemOpen
  \bibfield  {author} {\bibinfo {author} {\bibfnamefont {S.}~\bibnamefont
  {Lee}}, \bibinfo {author} {\bibfnamefont {S.}~\bibnamefont {Trebst}},
  \bibinfo {author} {\bibfnamefont {Y.~B.}\ \bibnamefont {Kim}}, \ and\
  \bibinfo {author} {\bibfnamefont {A.}~\bibnamefont {Paramekanti}},\ }\href
  {\doibase 10.1103/PhysRevB.98.134447} {\bibfield  {journal} {\bibinfo
  {journal} {Phys. Rev. B}\ }\textbf {\bibinfo {volume} {98}},\ \bibinfo
  {pages} {134447} (\bibinfo {year} {2018})}\BibitemShut {NoStop}%
\bibitem [{\citenamefont {Si}\ and\ \citenamefont {Smith}(1996)}]{Si1996}%
  \BibitemOpen
  \bibfield  {author} {\bibinfo {author} {\bibfnamefont {Q.}~\bibnamefont
  {Si}}\ and\ \bibinfo {author} {\bibfnamefont {J.~L.}\ \bibnamefont {Smith}},\
  }\href {\doibase 10.1103/PhysRevLett.77.3391} {\bibfield  {journal} {\bibinfo
   {journal} {Physical Review Letters}\ }\textbf {\bibinfo {volume} {77}},\
  \bibinfo {pages} {3391} (\bibinfo {year} {1996})}\BibitemShut {NoStop}%
\bibitem [{\citenamefont {Smith}\ and\ \citenamefont {Si}(1999)}]{Smith1999}%
  \BibitemOpen
  \bibfield  {author} {\bibinfo {author} {\bibfnamefont {J.~L.}\ \bibnamefont
  {Smith}}\ and\ \bibinfo {author} {\bibfnamefont {Q.}~\bibnamefont {Si}},\
  }\href {\doibase 10.1209/epl/i1999-00151-4} {\bibfield  {journal} {\bibinfo
  {journal} {Europhysics Letters (EPL)}\ }\textbf {\bibinfo {volume} {45}},\
  \bibinfo {pages} {228} (\bibinfo {year} {1999})}\BibitemShut {NoStop}%
\bibitem [{\citenamefont {Smith}\ and\ \citenamefont {Si}(2000)}]{Smith2000}%
  \BibitemOpen
  \bibfield  {author} {\bibinfo {author} {\bibfnamefont {J.~L.}\ \bibnamefont
  {Smith}}\ and\ \bibinfo {author} {\bibfnamefont {Q.}~\bibnamefont {Si}},\
  }\href {\doibase 10.1103/PhysRevB.61.5184} {\bibfield  {journal} {\bibinfo
  {journal} {Physical Review B - Condensed Matter and Materials Physics}\
  }\textbf {\bibinfo {volume} {61}},\ \bibinfo {pages} {5184} (\bibinfo {year}
  {2000})}\BibitemShut {NoStop}%
\bibitem [{\citenamefont {Si}\ \emph {et~al.}(2001)\citenamefont {Si},
  \citenamefont {Rabello}, \citenamefont {Ingersent},\ and\ \citenamefont
  {Smith}}]{Si2001}%
  \BibitemOpen
  \bibfield  {author} {\bibinfo {author} {\bibfnamefont {Q.}~\bibnamefont
  {Si}}, \bibinfo {author} {\bibfnamefont {S.}~\bibnamefont {Rabello}},
  \bibinfo {author} {\bibfnamefont {K.}~\bibnamefont {Ingersent}}, \ and\
  \bibinfo {author} {\bibfnamefont {J.~L.}\ \bibnamefont {Smith}},\ }\href
  {\doibase 10.1038/35101507} {\bibfield  {journal} {\bibinfo  {journal}
  {Nature}\ }\textbf {\bibinfo {volume} {413}},\ \bibinfo {pages} {804}
  (\bibinfo {year} {2001})}\BibitemShut {NoStop}%
\bibitem [{\citenamefont {Si}\ \emph {et~al.}(2003)\citenamefont {Si},
  \citenamefont {Rabello}, \citenamefont {Ingersent},\ and\ \citenamefont
  {Smith}}]{Si2003}%
  \BibitemOpen
  \bibfield  {author} {\bibinfo {author} {\bibfnamefont {Q.}~\bibnamefont
  {Si}}, \bibinfo {author} {\bibfnamefont {S.}~\bibnamefont {Rabello}},
  \bibinfo {author} {\bibfnamefont {K.}~\bibnamefont {Ingersent}}, \ and\
  \bibinfo {author} {\bibfnamefont {J.~L.}\ \bibnamefont {Smith}},\ }\href
  {\doibase 10.1103/PhysRevB.68.115103} {\bibfield  {journal} {\bibinfo
  {journal} {Phys. Rev. B}\ }\textbf {\bibinfo {volume} {68}},\ \bibinfo
  {pages} {115103} (\bibinfo {year} {2003})}\BibitemShut {NoStop}%
\bibitem [{\citenamefont {Sengupta}(2000)}]{Sengupta2000}%
  \BibitemOpen
  \bibfield  {author} {\bibinfo {author} {\bibfnamefont {A.~M.}\ \bibnamefont
  {Sengupta}},\ }\href {\doibase 10.1103/PhysRevB.61.4041} {\bibfield
  {journal} {\bibinfo  {journal} {Physical Review B}\ }\textbf {\bibinfo
  {volume} {61}},\ \bibinfo {pages} {4041} (\bibinfo {year}
  {2000})}\BibitemShut {NoStop}%
\bibitem [{\citenamefont {Zar\'and}\ and\ \citenamefont
  {Demler}(2002)}]{Zarand2002}%
  \BibitemOpen
  \bibfield  {author} {\bibinfo {author} {\bibfnamefont {G.}~\bibnamefont
  {Zar\'and}}\ and\ \bibinfo {author} {\bibfnamefont {E.}~\bibnamefont
  {Demler}},\ }\href {\doibase 10.1103/PhysRevB.66.024427} {\bibfield
  {journal} {\bibinfo  {journal} {Phys. Rev. B}\ }\textbf {\bibinfo {volume}
  {66}},\ \bibinfo {pages} {024427} (\bibinfo {year} {2002})}\BibitemShut
  {NoStop}%
\bibitem [{\citenamefont {Kirchner}\ \emph {et~al.}(2005)\citenamefont
  {Kirchner}, \citenamefont {Zhu},\ and\ \citenamefont {Si}}]{Kirchner2005}%
  \BibitemOpen
  \bibfield  {author} {\bibinfo {author} {\bibfnamefont {S.}~\bibnamefont
  {Kirchner}}, \bibinfo {author} {\bibfnamefont {L.}~\bibnamefont {Zhu}}, \
  and\ \bibinfo {author} {\bibfnamefont {Q.}~\bibnamefont {Si}},\ }\href
  {\doibase 10.1016/j.physb.2004.12.064} {\bibfield  {journal} {\bibinfo
  {journal} {Physica B: Condensed Matter}\ }\textbf {\bibinfo {volume}
  {359-361}},\ \bibinfo {pages} {83} (\bibinfo {year} {2005})}\BibitemShut
  {NoStop}%
\bibitem [{\citenamefont {Zhu}\ and\ \citenamefont {Si}(2002)}]{Zhu2002}%
  \BibitemOpen
  \bibfield  {author} {\bibinfo {author} {\bibfnamefont {L.}~\bibnamefont
  {Zhu}}\ and\ \bibinfo {author} {\bibfnamefont {Q.}~\bibnamefont {Si}},\
  }\href {\doibase 10.1103/PhysRevB.66.024426} {\bibfield  {journal} {\bibinfo
  {journal} {Phys. Rev. B}\ }\textbf {\bibinfo {volume} {66}},\ \bibinfo
  {pages} {024426} (\bibinfo {year} {2002})}\BibitemShut {NoStop}%
\bibitem [{\citenamefont {Han}\ and\ \citenamefont {Yu}(2021)}]{Han2021}%
  \BibitemOpen
  \bibfield  {author} {\bibinfo {author} {\bibfnamefont {X.}~\bibnamefont
  {Han}}\ and\ \bibinfo {author} {\bibfnamefont {Z.}~\bibnamefont {Yu}},\
  }\href {\doibase 10.1103/PhysRevB.104.085139} {\bibfield  {journal} {\bibinfo
   {journal} {Phys. Rev. B}\ }\textbf {\bibinfo {volume} {104}},\ \bibinfo
  {pages} {085139} (\bibinfo {year} {2021})}\BibitemShut {NoStop}%
\bibitem [{\citenamefont {Steglich}\ \emph {et~al.}(2014)\citenamefont
  {Steglich}, \citenamefont {Pfau}, \citenamefont {Lausberg}, \citenamefont
  {Hamann}, \citenamefont {Sun}, \citenamefont {Stockert}, \citenamefont
  {Brando}, \citenamefont {Friedemann}, \citenamefont {Krellner}, \citenamefont
  {Geibel}, \citenamefont {Wirth}, \citenamefont {Kirchner}, \citenamefont
  {Abrahams},\ and\ \citenamefont {Si}}]{Steglich2014}%
  \BibitemOpen
  \bibfield  {author} {\bibinfo {author} {\bibfnamefont {F.}~\bibnamefont
  {Steglich}}, \bibinfo {author} {\bibfnamefont {H.}~\bibnamefont {Pfau}},
  \bibinfo {author} {\bibfnamefont {S.}~\bibnamefont {Lausberg}}, \bibinfo
  {author} {\bibfnamefont {S.}~\bibnamefont {Hamann}}, \bibinfo {author}
  {\bibfnamefont {P.}~\bibnamefont {Sun}}, \bibinfo {author} {\bibfnamefont
  {U.}~\bibnamefont {Stockert}}, \bibinfo {author} {\bibfnamefont
  {M.}~\bibnamefont {Brando}}, \bibinfo {author} {\bibfnamefont
  {S.}~\bibnamefont {Friedemann}}, \bibinfo {author} {\bibfnamefont
  {C.}~\bibnamefont {Krellner}}, \bibinfo {author} {\bibfnamefont
  {C.}~\bibnamefont {Geibel}}, \bibinfo {author} {\bibfnamefont
  {S.}~\bibnamefont {Wirth}}, \bibinfo {author} {\bibfnamefont
  {S.}~\bibnamefont {Kirchner}}, \bibinfo {author} {\bibfnamefont
  {E.}~\bibnamefont {Abrahams}}, \ and\ \bibinfo {author} {\bibfnamefont
  {Q.}~\bibnamefont {Si}},\ }\href {\doibase 10.7566/JPSJ.83.061001} {\bibfield
   {journal} {\bibinfo  {journal} {Journal of the Physical Society of Japan}\
  }\textbf {\bibinfo {volume} {83}},\ \bibinfo {pages} {061001} (\bibinfo
  {year} {2014})}\BibitemShut {NoStop}%
\bibitem [{\citenamefont {Friedemann}\ \emph
  {et~al.}(2010{\natexlab{a}})\citenamefont {Friedemann}, \citenamefont
  {Wirth}, \citenamefont {Oeschler}, \citenamefont {Krellner}, \citenamefont
  {Geibel}, \citenamefont {Steglich}, \citenamefont {MaQuilon}, \citenamefont
  {Fisk}, \citenamefont {Paschen},\ and\ \citenamefont
  {Zwicknagl}}]{Friedemann2010}%
  \BibitemOpen
  \bibfield  {author} {\bibinfo {author} {\bibfnamefont {S.}~\bibnamefont
  {Friedemann}}, \bibinfo {author} {\bibfnamefont {S.}~\bibnamefont {Wirth}},
  \bibinfo {author} {\bibfnamefont {N.}~\bibnamefont {Oeschler}}, \bibinfo
  {author} {\bibfnamefont {C.}~\bibnamefont {Krellner}}, \bibinfo {author}
  {\bibfnamefont {C.}~\bibnamefont {Geibel}}, \bibinfo {author} {\bibfnamefont
  {F.}~\bibnamefont {Steglich}}, \bibinfo {author} {\bibfnamefont
  {S.}~\bibnamefont {MaQuilon}}, \bibinfo {author} {\bibfnamefont
  {Z.}~\bibnamefont {Fisk}}, \bibinfo {author} {\bibfnamefont {S.}~\bibnamefont
  {Paschen}}, \ and\ \bibinfo {author} {\bibfnamefont {G.}~\bibnamefont
  {Zwicknagl}},\ }\href {\doibase 10.1103/PhysRevB.82.035103} {\bibfield
  {journal} {\bibinfo  {journal} {Phys. Rev. B}\ }\textbf {\bibinfo {volume}
  {82}},\ \bibinfo {pages} {035103} (\bibinfo {year}
  {2010}{\natexlab{a}})}\BibitemShut {NoStop}%
\bibitem [{\citenamefont {Friedemann}\ \emph
  {et~al.}(2010{\natexlab{b}})\citenamefont {Friedemann}, \citenamefont
  {Oeschler}, \citenamefont {Wirth}, \citenamefont {Krellner}, \citenamefont
  {Geibel}, \citenamefont {Steglich}, \citenamefont {Paschen}, \citenamefont
  {Kirchner},\ and\ \citenamefont {Si}}]{Friedemann2010a}%
  \BibitemOpen
  \bibfield  {author} {\bibinfo {author} {\bibfnamefont {S.}~\bibnamefont
  {Friedemann}}, \bibinfo {author} {\bibfnamefont {N.}~\bibnamefont
  {Oeschler}}, \bibinfo {author} {\bibfnamefont {S.}~\bibnamefont {Wirth}},
  \bibinfo {author} {\bibfnamefont {C.}~\bibnamefont {Krellner}}, \bibinfo
  {author} {\bibfnamefont {C.}~\bibnamefont {Geibel}}, \bibinfo {author}
  {\bibfnamefont {F.}~\bibnamefont {Steglich}}, \bibinfo {author}
  {\bibfnamefont {S.}~\bibnamefont {Paschen}}, \bibinfo {author} {\bibfnamefont
  {S.}~\bibnamefont {Kirchner}}, \ and\ \bibinfo {author} {\bibfnamefont
  {Q.}~\bibnamefont {Si}},\ }\href {\doibase 10.1073/pnas.1009202107}
  {\bibfield  {journal} {\bibinfo  {journal} {Proceedings of the National
  Academy of Sciences of the United States of America}\ }\textbf {\bibinfo
  {volume} {107}},\ \bibinfo {pages} {14547} (\bibinfo {year}
  {2010}{\natexlab{b}})}\BibitemShut {NoStop}%
\bibitem [{\citenamefont {Paschen}\ \emph {et~al.}(2004)\citenamefont
  {Paschen}, \citenamefont {L{\"{u}}hmann}, \citenamefont {Wirth},
  \citenamefont {Gegenwart}, \citenamefont {Trovarelli}, \citenamefont
  {Geibel}, \citenamefont {Steglich}, \citenamefont {Coleman},\ and\
  \citenamefont {Si}}]{Paschen2004}%
  \BibitemOpen
  \bibfield  {author} {\bibinfo {author} {\bibfnamefont {S.}~\bibnamefont
  {Paschen}}, \bibinfo {author} {\bibfnamefont {T.}~\bibnamefont
  {L{\"{u}}hmann}}, \bibinfo {author} {\bibfnamefont {S.}~\bibnamefont
  {Wirth}}, \bibinfo {author} {\bibfnamefont {P.}~\bibnamefont {Gegenwart}},
  \bibinfo {author} {\bibfnamefont {O.}~\bibnamefont {Trovarelli}}, \bibinfo
  {author} {\bibfnamefont {C.}~\bibnamefont {Geibel}}, \bibinfo {author}
  {\bibfnamefont {F.}~\bibnamefont {Steglich}}, \bibinfo {author}
  {\bibfnamefont {P.}~\bibnamefont {Coleman}}, \ and\ \bibinfo {author}
  {\bibfnamefont {Q.}~\bibnamefont {Si}},\ }\href {\doibase 10.1038/nature03129} {\bibfield  {journal} {\bibinfo  {journal} {Nature}\
  }\textbf {\bibinfo {volume} {432}},\ \bibinfo {pages} {881} (\bibinfo {year}
  {2004})}\BibitemShut {NoStop}%
\bibitem [{\citenamefont {Coleman}\ \emph {et~al.}(2001)\citenamefont
  {Coleman}, \citenamefont {P{\'{e}}pin}, \citenamefont {Si},\ and\
  \citenamefont {Ramazashvili}}]{Coleman2001}%
  \BibitemOpen
  \bibfield  {author} {\bibinfo {author} {\bibfnamefont {P.}~\bibnamefont
  {Coleman}}, \bibinfo {author} {\bibfnamefont {C.}~\bibnamefont
  {P{\'{e}}pin}}, \bibinfo {author} {\bibfnamefont {Q.}~\bibnamefont {Si}}, \
  and\ \bibinfo {author} {\bibfnamefont {R.}~\bibnamefont {Ramazashvili}},\
  }\href {\doibase 10.1088/0953-8984/13/35/202} {\bibfield  {journal} {\bibinfo
   {journal} {Journal of Physics Condensed Matter}\ }\textbf {\bibinfo {volume}
  {13}},\ \bibinfo {pages} {723} (\bibinfo {year} {2001})}\BibitemShut
  {NoStop}%
\bibitem [{\citenamefont {L{\"{o}}hneysen}\ \emph {et~al.}(2007)\citenamefont
  {L{\"{o}}hneysen}, \citenamefont {Rosch}, \citenamefont {Vojta},\ and\
  \citenamefont {W{\"{o}}lfle}}]{Lohneysen2007}%
  \BibitemOpen
  \bibfield  {author} {\bibinfo {author} {\bibfnamefont {H.~V.}\ \bibnamefont
  {L{\"{o}}hneysen}}, \bibinfo {author} {\bibfnamefont {A.}~\bibnamefont
  {Rosch}}, \bibinfo {author} {\bibfnamefont {M.}~\bibnamefont {Vojta}}, \ and\
  \bibinfo {author} {\bibfnamefont {P.}~\bibnamefont {W{\"{o}}lfle}},\ }\href
  {\doibase 10.1103/RevModPhys.79.1015} {\bibfield  {journal} {\bibinfo
  {journal} {Reviews of Modern Physics}\ }\textbf {\bibinfo {volume} {79}},\
  \bibinfo {pages} {1015} (\bibinfo {year} {2007})}\BibitemShut {NoStop}%
\bibitem [{\citenamefont {Glossop}\ and\ \citenamefont
  {Ingersent}(2007)}]{Glossop2007}%
  \BibitemOpen
  \bibfield  {author} {\bibinfo {author} {\bibfnamefont {M.~T.}\ \bibnamefont
  {Glossop}}\ and\ \bibinfo {author} {\bibfnamefont {K.}~\bibnamefont
  {Ingersent}},\ }\href {\doibase 10.1103/PhysRevLett.99.227203} {\bibfield
  {journal} {\bibinfo  {journal} {Phys. Rev. Lett.}\ }\textbf {\bibinfo
  {volume} {99}},\ \bibinfo {pages} {227203} (\bibinfo {year}
  {2007})}\BibitemShut {NoStop}%
\bibitem [{\citenamefont {Zhu}\ \emph {et~al.}(2007)\citenamefont {Zhu},
  \citenamefont {Kirchner}, \citenamefont {Bulla},\ and\ \citenamefont
  {Si}}]{Zhu2007}%
  \BibitemOpen
  \bibfield  {author} {\bibinfo {author} {\bibfnamefont {J.-X.}\ \bibnamefont
  {Zhu}}, \bibinfo {author} {\bibfnamefont {S.}~\bibnamefont {Kirchner}},
  \bibinfo {author} {\bibfnamefont {R.}~\bibnamefont {Bulla}}, \ and\ \bibinfo
  {author} {\bibfnamefont {Q.}~\bibnamefont {Si}},\ }\href {\doibase 10.1103/PhysRevLett.99.227204} {\bibfield  {journal} {\bibinfo  {journal}
  {Phys. Rev. Lett.}\ }\textbf {\bibinfo {volume} {99}},\ \bibinfo {pages}
  {227204} (\bibinfo {year} {2007})}\BibitemShut {NoStop}%
\bibitem [{\citenamefont {Yamane}\ \emph {et~al.}(2018)\citenamefont {Yamane},
  \citenamefont {Onimaru}, \citenamefont {Wakiya}, \citenamefont {Matsumoto},
  \citenamefont {Umeo},\ and\ \citenamefont {Takabatake}}]{Yamane2018b}%
  \BibitemOpen
  \bibfield  {author} {\bibinfo {author} {\bibfnamefont {Y.}~\bibnamefont
  {Yamane}}, \bibinfo {author} {\bibfnamefont {T.}~\bibnamefont {Onimaru}},
  \bibinfo {author} {\bibfnamefont {K.}~\bibnamefont {Wakiya}}, \bibinfo
  {author} {\bibfnamefont {K.~T.}\ \bibnamefont {Matsumoto}}, \bibinfo {author}
  {\bibfnamefont {K.}~\bibnamefont {Umeo}}, \ and\ \bibinfo {author}
  {\bibfnamefont {T.}~\bibnamefont {Takabatake}},\ }\href {\doibase 10.1103/PhysRevLett.121.077206} {\bibfield  {journal} {\bibinfo  {journal}
  {Physical Review Letters}\ }\textbf {\bibinfo {volume} {121}},\ \bibinfo
  {pages} {077206} (\bibinfo {year} {2018})}\BibitemShut {NoStop}%
\bibitem [{\citenamefont {Yanagisawa}\ \emph {et~al.}(2019)\citenamefont
  {Yanagisawa}, \citenamefont {Hidaka}, \citenamefont {Amitsuka}, \citenamefont
  {Zherlitsyn}, \citenamefont {Wosnitza}, \citenamefont {Yamane},\ and\
  \citenamefont {Onimaru}}]{Yanagisawa2019b}%
  \BibitemOpen
  \bibfield  {author} {\bibinfo {author} {\bibfnamefont {T.}~\bibnamefont
  {Yanagisawa}}, \bibinfo {author} {\bibfnamefont {H.}~\bibnamefont {Hidaka}},
  \bibinfo {author} {\bibfnamefont {H.}~\bibnamefont {Amitsuka}}, \bibinfo
  {author} {\bibfnamefont {S.}~\bibnamefont {Zherlitsyn}}, \bibinfo {author}
  {\bibfnamefont {J.}~\bibnamefont {Wosnitza}}, \bibinfo {author}
  {\bibfnamefont {Y.}~\bibnamefont {Yamane}}, \ and\ \bibinfo {author}
  {\bibfnamefont {T.}~\bibnamefont {Onimaru}},\ }\href {\doibase 10.1103/PhysRevLett.123.067201} {\bibfield  {journal} {\bibinfo  {journal}
  {Phys. Rev. Lett.}\ }\textbf {\bibinfo {volume} {123}},\ \bibinfo {pages}
  {067201} (\bibinfo {year} {2019})}\BibitemShut {NoStop}%
\bibitem [{\citenamefont {Yanagisawa}\ \emph {et~al.}()\citenamefont
  {Yanagisawa}, \citenamefont {Hidaka}, \citenamefont {Amitsuka}, \citenamefont
  {Zherlitsyn}, \citenamefont {Wosnitza}, \citenamefont {Yamane},\ and\
  \citenamefont {Onimaru}}]{Yanagisawa2019c}%
  \BibitemOpen
  \bibfield  {author} {\bibinfo {author} {\bibfnamefont {T.}~\bibnamefont
  {Yanagisawa}}, \bibinfo {author} {\bibfnamefont {H.}~\bibnamefont {Hidaka}},
  \bibinfo {author} {\bibfnamefont {H.}~\bibnamefont {Amitsuka}}, \bibinfo
  {author} {\bibfnamefont {S.}~\bibnamefont {Zherlitsyn}}, \bibinfo {author}
  {\bibfnamefont {J.}~\bibnamefont {Wosnitza}}, \bibinfo {author}
  {\bibfnamefont {Y.}~\bibnamefont {Yamane}}, \ and\ \bibinfo {author}
  {\bibfnamefont {T.}~\bibnamefont {Onimaru}},\ }in\ \href {\doibase 10.7566/JPSCP.29.015002} {\emph {\bibinfo {booktitle} {Proceedings of
  J-Physics 2019: International Conference on Multipole Physics and Related
  Phenomena}}}\BibitemShut {NoStop}%
\bibitem [{\citenamefont {Abrikosov}(1965)}]{Abrikosov1965}%
  \BibitemOpen
  \bibfield  {author} {\bibinfo {author} {\bibfnamefont {A.~A.}\ \bibnamefont
  {Abrikosov}},\ }\href {\doibase 10.1103/PhysicsPhysiqueFizika.2.5} {\bibfield
   {journal} {\bibinfo  {journal} {Physics Physique Fizika}\ }\textbf {\bibinfo
  {volume} {2}},\ \bibinfo {pages} {5} (\bibinfo {year} {1965})}\BibitemShut
  {NoStop}%
\bibitem [{\citenamefont {Ruderman}\ and\ \citenamefont
  {Kittel}(1954)}]{Ruderman1954a}%
  \BibitemOpen
  \bibfield  {author} {\bibinfo {author} {\bibfnamefont {M.~A.}\ \bibnamefont
  {Ruderman}}\ and\ \bibinfo {author} {\bibfnamefont {C.}~\bibnamefont
  {Kittel}},\ }\href {\doibase 10.1103/PhysRev.96.99} {\bibfield  {journal}
  {\bibinfo  {journal} {Physical Review}\ }\textbf {\bibinfo {volume} {96}},\
  \bibinfo {pages} {99} (\bibinfo {year} {1954})}\BibitemShut {NoStop}%
\bibitem [{\citenamefont {Kasuya}(1956)}]{Kasuya1956}%
  \BibitemOpen
  \bibfield  {author} {\bibinfo {author} {\bibfnamefont {T.}~\bibnamefont
  {Kasuya}},\ }\href {\doibase 10.1143/ptp.16.45} {\bibfield  {journal}
  {\bibinfo  {journal} {Progress of Theoretical Physics}\ }\textbf {\bibinfo
  {volume} {16}},\ \bibinfo {pages} {45} (\bibinfo {year} {1956})}\BibitemShut
  {NoStop}%
\bibitem [{\citenamefont {Yosida}(1957)}]{Yosida1957}%
  \BibitemOpen
  \bibfield  {author} {\bibinfo {author} {\bibfnamefont {K.}~\bibnamefont
  {Yosida}},\ }\href {\doibase 10.1103/PhysRev.106.893} {\bibfield  {journal}
  {\bibinfo  {journal} {Physical Review}\ }\textbf {\bibinfo {volume} {106}},\
  \bibinfo {pages} {893} (\bibinfo {year} {1957})}\BibitemShut {NoStop}%
\bibitem [{\citenamefont {Affleck}\ and\ \citenamefont
  {Ludwig}(1993)}]{Affleck1993b}%
  \BibitemOpen
  \bibfield  {author} {\bibinfo {author} {\bibfnamefont {I.}~\bibnamefont
  {Affleck}}\ and\ \bibinfo {author} {\bibfnamefont {A.~W.~W.}\ \bibnamefont
  {Ludwig}},\ }\href {\doibase 10.1103/PhysRevB.48.7297} {\bibfield  {journal}
  {\bibinfo  {journal} {Physical Review B}\ }\textbf {\bibinfo {volume} {48}},\
  \bibinfo {pages} {7297} (\bibinfo {year} {1993})}\BibitemShut {NoStop}%
\bibitem [{\citenamefont {Aronson}\ \emph {et~al.}(1997)\citenamefont
  {Aronson}, \citenamefont {Maple}, \citenamefont {de~Sa}, \citenamefont
  {Tsvelik},\ and\ \citenamefont {Osborn}}]{Aronson1997}%
  \BibitemOpen
  \bibfield  {author} {\bibinfo {author} {\bibfnamefont {M.~C.}\ \bibnamefont
  {Aronson}}, \bibinfo {author} {\bibfnamefont {M.~B.}\ \bibnamefont {Maple}},
  \bibinfo {author} {\bibfnamefont {P.}~\bibnamefont {de~Sa}}, \bibinfo
  {author} {\bibfnamefont {A.~M.}\ \bibnamefont {Tsvelik}}, \ and\ \bibinfo
  {author} {\bibfnamefont {R.}~\bibnamefont {Osborn}},\ }\href {\doibase 10.1209/epl/i1997-00455-9} {\bibfield  {journal} {\bibinfo  {journal}
  {Europhysics Letters (EPL)}\ }\textbf {\bibinfo {volume} {40}},\ \bibinfo
  {pages} {245} (\bibinfo {year} {1997})}\BibitemShut {NoStop}%
\bibitem [{\citenamefont {Yanagisawa}\ \emph {et~al.}(2018)\citenamefont
  {Yanagisawa}, \citenamefont {Mombetsu}, \citenamefont {Hidaka}, \citenamefont
  {Amitsuka}, \citenamefont {Cong}, \citenamefont {Yasin}, \citenamefont
  {Zherlitsyn}, \citenamefont {Wosnitza}, \citenamefont {Huang}, \citenamefont
  {Kanchanavatee}, \citenamefont {Janoschek}, \citenamefont {Maple},\ and\
  \citenamefont {Aoki}}]{Yanagisawa2018}%
  \BibitemOpen
  \bibfield  {author} {\bibinfo {author} {\bibfnamefont {T.}~\bibnamefont
  {Yanagisawa}}, \bibinfo {author} {\bibfnamefont {S.}~\bibnamefont
  {Mombetsu}}, \bibinfo {author} {\bibfnamefont {H.}~\bibnamefont {Hidaka}},
  \bibinfo {author} {\bibfnamefont {H.}~\bibnamefont {Amitsuka}}, \bibinfo
  {author} {\bibfnamefont {P.~T.}\ \bibnamefont {Cong}}, \bibinfo {author}
  {\bibfnamefont {S.}~\bibnamefont {Yasin}}, \bibinfo {author} {\bibfnamefont
  {S.}~\bibnamefont {Zherlitsyn}}, \bibinfo {author} {\bibfnamefont
  {J.}~\bibnamefont {Wosnitza}}, \bibinfo {author} {\bibfnamefont
  {K.}~\bibnamefont {Huang}}, \bibinfo {author} {\bibfnamefont
  {N.}~\bibnamefont {Kanchanavatee}}, \bibinfo {author} {\bibfnamefont
  {M.}~\bibnamefont {Janoschek}}, \bibinfo {author} {\bibfnamefont {M.~B.}\
  \bibnamefont {Maple}}, \ and\ \bibinfo {author} {\bibfnamefont
  {D.}~\bibnamefont {Aoki}},\ }\href {\doibase 10.1103/PhysRevB.97.155137}
  {\bibfield  {journal} {\bibinfo  {journal} {Phys. Rev. B}\ }\textbf {\bibinfo
  {volume} {97}},\ \bibinfo {pages} {155137} (\bibinfo {year}
  {2018})}\BibitemShut {NoStop}%
\bibitem [{\citenamefont {Sorensen}\ and\ \citenamefont
  {Fisher}(2021)}]{Sorensen2021}%
  \BibitemOpen
  \bibfield  {author} {\bibinfo {author} {\bibfnamefont {M.~E.}\ \bibnamefont
  {Sorensen}}\ and\ \bibinfo {author} {\bibfnamefont {I.~R.}\ \bibnamefont
  {Fisher}},\ }\href {\doibase 10.1103/PhysRevB.103.155106} {\bibfield
  {journal} {\bibinfo  {journal} {Physical Review B}\ }\textbf {\bibinfo
  {volume} {103}},\ \bibinfo {pages} {155106} (\bibinfo {year}
  {2021})}\BibitemShut {NoStop}%
\bibitem [{\citenamefont {Tsuruta}\ \emph
  {et~al.}(2000{\natexlab{b}})\citenamefont {Tsuruta}, \citenamefont
  {Kobayashi}, \citenamefont {Matsuura},\ and\ \citenamefont
  {Kuroda}}]{Tsuruta2000}%
  \BibitemOpen
  \bibfield  {author} {\bibinfo {author} {\bibfnamefont {A.}~\bibnamefont
  {Tsuruta}}, \bibinfo {author} {\bibfnamefont {A.}~\bibnamefont {Kobayashi}},
  \bibinfo {author} {\bibfnamefont {T.}~\bibnamefont {Matsuura}}, \ and\
  \bibinfo {author} {\bibfnamefont {Y.}~\bibnamefont {Kuroda}},\ }\href
  {\doibase 10.1143/JPSJ.69.663} {\bibfield  {journal} {\bibinfo  {journal}
  {Journal of the Physical Society of Japan}\ }\textbf {\bibinfo {volume}
  {69}},\ \bibinfo {pages} {663} (\bibinfo {year}
  {2000}{\natexlab{b}})}\BibitemShut {NoStop}%
\bibitem [{\citenamefont {Nakamura}\ \emph {et~al.}(2019)\citenamefont
  {Nakamura}, \citenamefont {Kano},\ and\ \citenamefont
  {Ohara}}]{Nakamura2019}%
  \BibitemOpen
  \bibfield  {author} {\bibinfo {author} {\bibfnamefont {S.}~\bibnamefont
  {Nakamura}}, \bibinfo {author} {\bibfnamefont {T.}~\bibnamefont {Kano}}, \
  and\ \bibinfo {author} {\bibfnamefont {S.}~\bibnamefont {Ohara}},\ }\href
  {\doibase 10.7566/JPSJ.88.093705} {\bibfield  {journal} {\bibinfo  {journal}
  {Journal of the Physical Society of Japan}\ }\textbf {\bibinfo {volume}
  {88}},\ \bibinfo {pages} {5} (\bibinfo {year} {2019})}\BibitemShut {NoStop}%
\bibitem [{\citenamefont {Schultz}\ \emph
  {et~al.}(2021{\natexlab{b}})\citenamefont {Schultz}, \citenamefont {Patri},\
  and\ \citenamefont {Kim}}]{Schultz2021c}%
  \BibitemOpen
  \bibfield  {author} {\bibinfo {author} {\bibfnamefont {D.~J.}\ \bibnamefont
  {Schultz}}, \bibinfo {author} {\bibfnamefont {A.~S.}\ \bibnamefont {Patri}},
  \ and\ \bibinfo {author} {\bibfnamefont {Y.~B.}\ \bibnamefont {Kim}},\ }\href
  {\doibase 10.1103/PhysRevB.104.125144} {\bibfield  {journal} {\bibinfo
  {journal} {Physical Review B}\ }\textbf {\bibinfo {volume} {104}},\ \bibinfo
  {pages} {125144} (\bibinfo {year} {2021}{\natexlab{b}})}\BibitemShut
  {NoStop}%
\bibitem [{\citenamefont {Liu}\ \emph {et~al.}(2021)\citenamefont {Liu},
  \citenamefont {Paschen},\ and\ \citenamefont {Si}}]{Liu2021}%
  \BibitemOpen
  \bibfield  {author} {\bibinfo {author} {\bibfnamefont {C.-C.}\ \bibnamefont
  {Liu}}, \bibinfo {author} {\bibfnamefont {S.}~\bibnamefont {Paschen}}, \ and\
  \bibinfo {author} {\bibfnamefont {Q.}~\bibnamefont {Si}},\ }\href
  {http://arxiv.org/abs/2101.01087} {\ ,\ \bibinfo {pages} {1} (\bibinfo {year}
  {2021})},\ \Eprint {http://arxiv.org/abs/2101.01087} {arXiv:2101.01087}
  \BibitemShut {NoStop}%
\bibitem [{\citenamefont {Shiina}\ \emph {et~al.}(1998)\citenamefont {Shiina},
  \citenamefont {Sakai}, \citenamefont {Shiba},\ and\ \citenamefont
  {Thalmeier}}]{Shiina1998}%
  \BibitemOpen
  \bibfield  {author} {\bibinfo {author} {\bibfnamefont {R.}~\bibnamefont
  {Shiina}}, \bibinfo {author} {\bibfnamefont {O.}~\bibnamefont {Sakai}},
  \bibinfo {author} {\bibfnamefont {H.}~\bibnamefont {Shiba}}, \ and\ \bibinfo
  {author} {\bibfnamefont {P.}~\bibnamefont {Thalmeier}},\ }\href {\doibase 10.1143/JPSJ.67.941} {\bibfield  {journal} {\bibinfo  {journal} {Journal of
  the Physical Society of Japan}\ }\textbf {\bibinfo {volume} {67}},\ \bibinfo
  {pages} {941} (\bibinfo {year} {1998})}\BibitemShut {NoStop}%
\bibitem [{\citenamefont {Shiina}\ \emph {et~al.}(1997)\citenamefont {Shiina},
  \citenamefont {Shiba},\ and\ \citenamefont {Thalmeier}}]{Shiina1997}%
  \BibitemOpen
  \bibfield  {author} {\bibinfo {author} {\bibfnamefont {R.}~\bibnamefont
  {Shiina}}, \bibinfo {author} {\bibfnamefont {H.}~\bibnamefont {Shiba}}, \
  and\ \bibinfo {author} {\bibfnamefont {P.}~\bibnamefont {Thalmeier}},\ }\href
  {\doibase 10.1143/JPSJ.66.1741} {\bibfield  {journal} {\bibinfo  {journal}
  {Journal of the Physical Society of Japan}\ }\textbf {\bibinfo {volume}
  {66}},\ \bibinfo {pages} {1741} (\bibinfo {year} {1997})}\BibitemShut
  {NoStop}%
\bibitem [{\citenamefont {Parcollet}\ \emph {et~al.}(1998)\citenamefont
  {Parcollet}, \citenamefont {Georges}, \citenamefont {Kotliar},\ and\
  \citenamefont {Sengupta}}]{Parcollet1998}%
  \BibitemOpen
  \bibfield  {author} {\bibinfo {author} {\bibfnamefont {O.}~\bibnamefont
  {Parcollet}}, \bibinfo {author} {\bibfnamefont {A.}~\bibnamefont {Georges}},
  \bibinfo {author} {\bibfnamefont {G.}~\bibnamefont {Kotliar}}, \ and\
  \bibinfo {author} {\bibfnamefont {A.}~\bibnamefont {Sengupta}},\ }\href
  {\doibase 10.1103/PhysRevB.58.3794} {\bibfield  {journal} {\bibinfo
  {journal} {Physical Review B}\ }\textbf {\bibinfo {volume} {58}},\ \bibinfo
  {pages} {3794} (\bibinfo {year} {1998})}\BibitemShut {NoStop}%
\bibitem [{\citenamefont {Parcollet}\ and\ \citenamefont
  {Georges}(1999)}]{Parcollet1999}%
  \BibitemOpen
  \bibfield  {author} {\bibinfo {author} {\bibfnamefont {O.}~\bibnamefont
  {Parcollet}}\ and\ \bibinfo {author} {\bibfnamefont {A.}~\bibnamefont
  {Georges}},\ }\href {\doibase 10.1103/PhysRevB.59.5341} {\bibfield  {journal}
  {\bibinfo  {journal} {Physical Review B}\ }\textbf {\bibinfo {volume} {59}},\
  \bibinfo {pages} {5341} (\bibinfo {year} {1999})}\BibitemShut {NoStop}%
\end{thebibliography}
%

\end{document}